\documentclass[a4paper,11pt,reqno]{article}
\usepackage{a4wide}

\usepackage{amsmath,amsfonts,amssymb}
\usepackage[T1]{fontenc}
\usepackage[p,osf]{cochineal}
\usepackage[varqu,varl,var0]{inconsolata}
\usepackage[scale=.95,type1]{cabin}
\usepackage[vvarbb]{newtxmath}
\usepackage[cal=boondoxo]{mathalfa}
\usepackage[mathscr]{euscript}
\usepackage{bbold} 
\usepackage{cite}
\usepackage[toc,page]{appendix}
\usepackage{graphicx}
\usepackage{musicography}
 \usepackage{soul}

\usepackage{lettrine} 
\usepackage{pict2e}

\makeatletter
\DeclareRobustCommand{\loplus}{\mathbin{\mathpalette\dog@lsemi{+}}}
\DeclareRobustCommand{\lotimes}{\mathbin{\mathpalette\dog@lsemi{\times}}}
\DeclareRobustCommand{\roplus}{\mathbin{\mathpalette\dog@rsemi{+}}}
\DeclareRobustCommand{\rotimes}{\mathbin{\mathpalette\dog@rsemi{\times}}}

\newcommand{\dog@rsemi}[2]{\dog@semi{#1}{#2}{-90,90}}
\newcommand{\dog@lsemi}[2]{\dog@semi{#1}{#2}{270,90}}
\newcommand{\dog@semi}[3]{%
  \begingroup
  \sbox\z@{$\m@th#1#2$}%
  \setlength{\unitlength}{\dimexpr\ht\z@+\dp\z@\relax}%
  \makebox[\wd\z@]{\raisebox{-\dp\z@}{%
    \begin{picture}(1,1)
    \linethickness{\variable@rule{#1}}
    \roundcap
    \put(0.5,0.5){\makebox(0,0){\raisebox{\dp\z@}{$\m@th#1#2$}}}
    \put(0.5,0.5){\arc[#3]{0.5}}
    \end{picture}%
  }}%
  \endgroup
}
\newcommand{\variable@rule}[1]{%
  \fontdimen8  
  \ifx#1\displaystyle\textfont3\else
    \ifx#1\textstyle\textfont3\else
      \ifx#1\scriptstyle\scriptfont3\else
        \scriptscriptfont3\relax
  \fi\fi\fi
}
\makeatother

\newcommand{\p}{\partial}
\newcommand{\scri}{{\cal I}}
\newcommand{\nn}{\nonumber}

\DeclareMathOperator{\tr}{tr}
\renewcommand{\textbf}[1]{\begingroup\bfseries\mathversion{bold}#1\endgroup}

 \usepackage{braket}

\usepackage{pict2e}

\newcommand{\be}{\begin{equation}}
\newcommand{\ee}{\end{equation}}
\newcommand{\barray}{\begin{array}}
\newcommand{\earray}{\end{array}}
\newcommand{\bea}{\begin{eqnarray}}
\newcommand{\eea}{\end{eqnarray}}
\newcommand{\bs}{\begin{subequations}}
\newcommand{\es}{\end{subequations}}

\newcommand{\bit}{\begin{itemize}}
\newcommand{\eit}{\end{itemize}}
\newcommand{\bd}{\begin{description}}
\newcommand{\ed}{\end{description}}

\def\w{\wedge}

\DeclareMathOperator{\Tr}{Tr}
\newcommand{\f}{\frac}

\renewcommand{\a}{\alpha}  \newcommand{\g}{\gamma}  
  \newcommand{\eps}{\epsilon}

\let\m=\mu        \let\om=\omega
 \newcommand{\s}{\sigma}      
\let\G=\Gamma     
 \let\Om=\Omega

\usepackage{nicefrac}
\usepackage{setspace}
\usepackage{datetime}

\usepackage{comment}

\usepackage[euler]{textgreek}
\usepackage[breaklinks=true]{hyperref}

\numberwithin{equation}{section}

\hypersetup{
			colorlinks=true,
			urlcolor=blue,
			citecolor=magenta,
			linkcolor=blue,
     }

\let\oldsqrt\sqrt
\def\sqrt{\mathpalette\DHLhksqrt}
\def\DHLhksqrt#1#2{%
\setbox0=\hbox{$#1\oldsqrt{#2\,}$}\dimen0=\ht0
\advance\dimen0-0.2\ht0
\setbox2=\hbox{\vrule height\ht0 depth -\dimen0}%
{\box0\lower0.4pt\box2}}

\newcommand{\RNum}[1]{\uppercase\expandafter{\romannumeral #1\relax}}

\author{
  \begin{minipage}{.97\linewidth}
    \vspace{1cm}
       \begin{center}
      \begin{small}
             \textbf{Antoine Rignon-Bret}$^{1,}$\footnote{\href{arignonbret@gmail.com}{arignonbret@gmail.com}} 
     \, and 
      \textbf{Matthieu Vilatte}$^{2,}$\footnote{\href{matthieu.vilatte@umons.ac.be}{matthieu.vilatte@umons.ac.be}}
              \end{small}
    \end{center}
    \vspace{0.5cm}
    \hspace{2.4cm}\begin{minipage}{.7\linewidth}
\begin{center}     {\it \begin{footnotesize}
\hbox{\kern-1.8cm\vbox{\vskip0cm
 \begin{itemize}
               \item[$^1$]Universit\'e de Lorraine, CNRS, \\ Laboratoire de Physique et de Chimie Th\'eoriques,\\
               F-54000, Nancy, France\\
                           \vskip0.25cm
      \end{itemize}}
\kern-3.2cm\vbox{
\begin{itemize}
 \item[$^2$]Service de Physique de l'Univers, Champs et Gravitation, \\
        Université de Mons -- UMONS, \\ 
        Place du Parc 20, 7000 Mons, Belgium         
      \end{itemize}
      \vskip0.cm
}}
     \end{footnotesize}}
\end{center}
    \end{minipage}
  \end{minipage}
}

\title{\vspace{1.5cm}
 \boldmath \begin{LARGE}
    \textbf{\textsc{Black hole thermodynamics at null infinity \\
    Part 2: Open systems, Markovian dynamics and work extraction from non-rotating black holes}}
  \end{LARGE} \unboldmath
}

\date{}

\begin{document}


\begin{titlepage}
\maketitle
\thispagestyle{empty}


\begin{center}
\textsc{Abstract}\\  
\vspace{1. cm}	
\begin{minipage}{1.0\linewidth}

Black hole thermodynamics provides a unique setting in which general relativity, quantum field theory, and statistical mechanics converge. In semiclassical gravity, this interplay culminates in the generalized second law (GSL), whose modern proofs rely on information-theoretic techniques applied to algebras of observables defined on null hypersurfaces. These proofs exhibit close structural parallels with the thermodynamics of open quantum systems governed by Markovian dynamics. In this work, we draw parallels between the dynamics of quantum fields in regions bounded by non-expanding causal horizons and the thermodynamics of quantum systems weakly coupled to equilibrium reservoirs. We introduce a dictionary relating late-time boundary conditions to the choice of reservoir, vacuum states to fixed points of the dynamics, and modular Hamiltonians to thermodynamic potentials. Building on results from a companion paper on dual generalized second laws at future null infinity \cite{RBVilatte251}, we show that additional terms appearing in the associated thermodynamic potentials admit a natural interpretation as work contributions. We demonstrate that certain non-thermal vacuum states at null infinity allow for the operation of autonomous thermal engines and enable work extraction from the radiation. Extending the analysis to the Unruh vacuum in Schwarzschild and Kerr backgrounds, we obtain generalized grand-potential–type laws incorporating grey-body effects and angular momentum fluxes. Altogether, our results clarify the thermodynamic description of black hole dynamics and place it within the broader framework of open quantum thermodynamics.

\end{minipage}
\end{center}

\end{titlepage}

\onehalfspace

\begingroup
\hypersetup{linkcolor=black}
\tableofcontents
\endgroup
\noindent\rule{\textwidth}{0.6pt}


\section{Introduction}
\label{sec: intro}

Black hole thermodynamics \cite{penrose1971extraction, Hawking:1971tu, hawking1972energy, bardeen1973four, bekenstein1973black, bekenstein1974generalized, Hawking:1974rv, hawking1975particle, Wald:1975kc, israel1976thermo, hartle1976path, Unruh:1976db} (see, for example, \cite{Wald:2002mon, Wall:2018ydq, Perez:2023ctt} for comprehensive reviews) has occupied a central role in research on quantum gravity for more than half a century. The subject raises a number of deep and interrelated challenges, including the precise formulation and derivation of the laws of black hole mechanics \cite{sudarsky1992extrema, wald1993black, iyer1994some, jacobson1994black, gao2001physical, prabhu2017first, Rignon-Bret:2023lyn}, their connection to Einstein's equations \cite{jacobson1995thermodynamics, jacobson2016entanglement, jacobson2019gravitational, Alonso-Serrano:2021uok, Alonso-Serrano:2025lbo} as well as the breakdown of predictability \cite{hawking1976breakdown, hawking1982unpredictability}, and proposed resolutions to the information loss problem \cite{page1993information, mathur2005fuzzball, Hawking:2015qqa, Hawking:2016msc, unruh2017information, Penington:2019npb, Penington:2019kki, Almheiri:2019psf, Almheiri:2019hni, Almheiri:2019qdq, Geng:2024xpj, Geng:2025efs, Bao:2025plr, Geng:2025rov, Geng:2025gns, Perez:2023ugg}. These issues continue to lie at the forefront of modern research in high-energy theoretical physics.

More recently, many efforts have been deployed to understand better the black hole entropy, either by extending the definition to the dynamical regime  \cite{chandrasekaran2018symmetries, Rignon-Bret:2023fjq, Odak:2023pga, Chandrasekaran:2023vzb, Ciambelli:2023mir, Hollands:2024vbe, Visser:2024pwz, Kong:2024sqc, Zhao:2025zny, Jia:2025tgf, Visser:2025jnf} or by using algebraic techniques to properly define the generalized entropy \cite{leutheusser2023causal, Witten:2021unn, Chandrasekaran:2022cip, Chandrasekaran:2022eqq, Jensen:2023yxy, Kudler-Flam:2023qfl, ali2024crossed, Ali:2024jkx, Faulkner:2024gst, DeVuyst:2024khu, DeVuyst:2024fxc}. The generalized entropy is tied to the generalized second law \cite{bekenstein1973black, bekenstein1974generalized, hawking1975particle}, according to which the total entropy of the universe, including the contribution from black holes, is non-decreasing. It can be written as
\be \label{genentropy}
    \Delta S_{\text{gen}} = \Delta \left(\frac{A}{4G} + S_{\text{out}} \right) \geq 0
\ee
where $A$ denotes the area of a cross section of the event horizon, and $S_{\text{out}}$ represents the entropy of the matter fields outside the black hole. Inspired by the work of Sorkin \cite{sorkin1998statistical} and Casini \cite{casini2008relative}, Wall, in a series of seminal papers \cite{Wall:2009wi, Wall:2010jtc, Wall:2010cj, Wall:2011hj}, successfully applied techniques from quantum information theory to quantum fields on the black hole horizon, providing a precise and rigorous formulation and derivation of \eqref{genentropy}. This framework has inspired a substantial body of subsequent works on the generalized second law \cite{Faulkner:2024gst, Wall:2015raa, Kapec:2016aqd, bousso2016generalized, hollands2022second, ARB24, Kirklin:2024gyl, ali2024crossed}. Moreover, the strategy developed in \cite{Wall:2011hj} relies on the monotonicity of the quantum relative entropy \cite{lindblad1975completely, uhlmann1977relative, araki1975relative, petz2003monotonicity, ohya2004quantum} and on the identification of an appropriate stationary reference state—techniques that are standard in the realm of thermodynamics of open quantum systems \cite{spohn1978entropy, schumacher2000relative, breuer2002theory, vedral2002role, goold2016role, alicki2019introduction, landi2021irreversible}. 

Hence, to better understand the generalized second law and, more broadly, the notion of gravitational entropy, it is instructive to compare proofs of the generalized second law on horizons with more standard proofs in non-relativistic settings. In particular, the theory of open quantum systems places strong emphasis on subsystems undergoing Markovian (i.e. memory-less) dynamics \cite{davies1974markovian, lindblad1976generators, spohn1978entropy, spohn1978irreversible, breuer2002theory, chruscinski2017brief, kosloff2019quantum, alicki2023thermodynamics}. The typical framework involves a small system weakly coupled to an effectively infinite bath at equilibrium (i.e. satisfying the Kubo-Martin-Schwinger (KMS)  conditions \cite{kubo1957statistical, martin1959theory}), with which it can exchange energy and particles. In this regime, the weak interaction induces transitions between the eigenstates of the small system. Moreover, when the coupling is sufficiently weak, correlations between the system and the environment can be neglected beyond a certain timescale, as they are dissipated into the infinite bath. This leads to the absence of information backflow and an effectively Markovian evolution \cite{breuer2002theory, manzano2020short}. A closely related structure appears in quantum field theory on null hypersurfaces. Since a horizon acts as a causal barrier, information that crosses it cannot return to the region from which it originated, and the dynamics of quantum fields in a causal region or on a null hypersurface is therefore Markovian. By contrast, when the boundary is timelike, radiation can re-enter the region, resulting in non-Markovian behavior. Moreover, whereas in open quantum systems the Markovian approximation is only valid to leading order, for quantum fields on a horizon it holds exactly, as a direct consequence of causality. 

Quantum thermodynamics, even when restricted to the Markovian regime, is an extremely rich subject \cite{breuer2002theory, brunner2012virtual, kosloff2013quantum, kosloff2014quantum, brandner2016periodic, hofer2017markovian, cresser2017coarse, liu2017heat, kosloff2019quantum, van2022thermodynamics}. Many physically relevant situations can be studied within this framework, such as relaxation processes towards (not necessarily thermal) stationary states or the operation of microscopic engines weakly coupled to reservoirs at different temperatures. A common feature of Markovian dynamics is that the system typically converges to a stationary state at late times. In quantum field theory, a natural analog of such a fixed point is a vacuum state, since both are characterized by invariance under an appropriate notion of time translation symmetry.
While in open quantum systems the fixed point of the dynamics is determined by the environment to which the system is coupled, quantum field theory instead exhibits a plurality of vacua associated with different representations of the algebra of observables. Of particular interest is the representation of observables defined on a black hole horizon or at null infinity using a vacuum state $\om$, which selects a set of modes with positive frequency with respect to a chosen notion of time on the null hypersurface.

The main goal of the present paper is to provide a detailed account of the correspondence between, on the one hand, the thermodynamics of a quantum system weakly coupled to an equilibrium reservoir and governed by Markovian dynamics, and, on the other hand, the thermodynamics induced on a non-expanding null hypersurface by a representation of the algebra of observables defined by a vacuum state compatible with prescribed late-time boundary conditions (especially in black hole backgrounds). Although we have already highlighted the close parallel between the proofs of the generalized second law and the thermodynamic laws emerging from Markovian dynamics, it is instructive to test these ideas in contexts beyond the standard formulation of the generalized second law. The present paper constitutes the second part of a two-part series, with the first part \cite{RBVilatte251} focusing on the precise setting in which the laws of spontaneous evolution are equivalent to the monotonicity of certain thermodynamic potentials constructed from observables at null infinity in a black hole background: these are the dual Generalized Second Laws.\footnote{The effective description of the thermodynamic potential was provided in \cite{ARB24}.} These proofs relied on selecting vacua invariant under specific symmetry transformations and on the monotonicity of the relative entropy. The thermodynamic potential appearing in the spontaneous evolution law was therefore tied to a particular choice of vacuum and, correspondingly, to a specific choice of Hilbert space.

In the present paper, we extend this analysis by identifying key quantities and concepts common to both frameworks (null hypersurfaces or open systems) and by establishing a precise dictionary between them. In particular, we clarify how the time scale identified in \cite{RBVilatte251}, which guarantees the monotonicity of the thermodynamic potentials, is related to the characteristic time scales that arise in Markovian evolution induced by an infinite equilibrium bath. More importantly, we show that the additional terms appearing in the thermodynamic potentials—when a more realistic description of the black hole radiation is considered—admit a natural interpretation as work contributions. Building on this perspective, we demonstrate how autonomous thermal engines can be operated using the $\kappa_l$-vacua introduced in \cite{RBVilatte251}, which provide a class of soft regularizations of the Hartle–Hawking vacuum. The underlying reason why this is possible is that these states are not thermal with respect to asymptotic observers. Consequently, they are not passive states \cite{pusz1978passive, alicki2019introduction}, and may instead be viewed as collections of reservoirs characterized by different effective temperatures, in close analogy with the description of the Hawking radiation spectrum in \cite{Page:2004xp}. This structure allows one to employ an autonomous engine—such as the Brunner–Linden–Popescu–Skrzypczyk engine \cite{brunner2012virtual}—to lift a load on an energy ladder, thereby extracting work from the non-thermal radiation.

The second part of the paper is devoted to completing and extending the results of \cite{RBVilatte251} to the Unruh vacuum, in both Schwarzschild and Kerr black hole backgrounds. In \cite{RBVilatte251}, quantum states were defined with respect to an asymptotic algebra of observables at null infinity. To ensure finite energy and entropy fluxes at null infinity, soft and hard regularizations of the Hartle–Hawking vacuum were introduced. The resulting class of vacuum states therefore interpolates between the Hartle–Hawking state and the more physically relevant Unruh vacuum, the latter providing an accurate description of Hawking's radiation at late times. However, it was emphasized in \cite{RBVilatte251} that the restriction of the Unruh vacuum itself to the asymptotic algebra of observables at null infinity does not belong to the class of regularized vacua considered there. Consequently, a natural and important objective is to obtain results analogous to those of \cite{RBVilatte251} directly for the Unruh vacuum.

Although an algebraic proof analogous to those obtained in \cite{RBVilatte251} for other classes of vacua is still lacking, we present several arguments supporting the expected form of the modular Hamiltonian of the Unruh vacuum when restricted to a subalgebra of $\scri^+$. These arguments rely on the symmetry properties of the Unruh vacuum on the past horizon \cite{kay1991theorems, dappiaggi2011rigorous, dappiaggi2017hadamard}, as well as on the contribution of the radiation transmitted through the potential barrier. Building on this structure, we are able to apply the same reasoning developed for the classes of vacua considered in \cite{RBVilatte251}. This allows us to express the dual generalized second law in terms of a grand potential whose chemical potentials are determined by the greybody factors. In this way, we recover results closely analogous to those obtained in \cite{ARB24}, which were derived using a more effective approach. 

\vspace{0.3cm}

The paper is organized as follows. Section~\ref{sec: GSL proof} presents a concise summary of the main results of \cite{RBVilatte251}, together with the key arguments underpinning them. We briefly review the semiclassical flux-balance laws on a perturbed Killing horizon and at null infinity, the quantization of a massless free scalar field on non-expanding null hypersurfaces, and the construction of algebras of observables on the Killing horizon and at null infinity. We also introduce the basic concepts of modular theory, review the definition of the relative entropy together with some of its fundamental properties, and recall the different classes of vacua introduced in \cite{RBVilatte251}. We discuss the symmetry transformations that leave these vacuum states invariant and explain how the monotonicity of the appropriate thermodynamic potential follows from these considerations.

In Section~\ref{sec: open systems}, we turn to conventional Markovian dynamics for open quantum systems. We introduce the Lindblad equation and discuss the approximations involved in its derivation, emphasizing how the assumptions required in the Lindblad framework closely parallel those made at null infinity in deriving the flux-balance laws. We then review the associated thermodynamics: the fixed points of the dynamics are determined by the KMS conditions of the reservoirs, which in general may not only exchange heat, but also work with the system. As a result, the stationary state does not need to be a thermal Gibbs state. We make the connection between quantum thermodynamics and the quantum-field-theoretic approach to the second law explicit and summarize the resulting parallels in a comparative table.

Section~\ref{sec: work extraction} extends this discussion by showing how the additional terms arising in the more intricate thermodynamic potentials at future null infinity can, in principle, be exploited to extract work and operate an engine, in close analogy with quantum open systems. To illustrate this mechanism, we employ the Brunner–Linden–Popescu–Skrzypczyk (BLPS) engine \cite{brunner2012virtual}, compute the maximum extractable work in the soft regularization of the Hartle-Hawking state using the Carnot universal efficiency bound, and precisely recover the chemical potentials introduced in \cite{RBVilatte251}.

Finally, in Section~\ref{sec: vac unruh}, we focus on the Unruh vacuum. Exploiting its symmetry properties on the past horizon, we derive a formal expression for the one-sided modular Hamiltonian associated with an algebra of observables at null infinity. Applying monotonicity arguments for the relative entropy yields a dual GSL statement closely analogous to that obtained for the $\kappa_l$-vacua in \cite{RBVilatte251}, albeit with modified expressions for the chemical potentials. We analyze the resulting thermodynamic potentials and their properties in parallel with Sections~\ref{sec: open systems} and~\ref{sec: work extraction}, for both Schwarzschild and Kerr black holes. In the Kerr case, an additional contribution appears, naturally interpreted as a work term proportional to the angular momentum flux.


\section{Dual Generalized Second Law}
\label{sec: GSL proof}

In this Section we summarize the results obtained in  \cite{RBVilatte251} regarding the dual Generalized Second Law. We start by reviewing some background notions such as the algebra of observables, its Hilbert space representations from the GNS construction, and the definition of the relative entropy. Then we recall the definition of maximally extended null infinity and the three vacuum states we considered in  \cite{RBVilatte251}. Finally we remind how one can get the dual GSL from the monotonicity of the relative entropy between a pair of nested algebra of observables at null infinity. We refer the reader to the aforementioned reference for more details.


\subsection{Theoretical background}
\label{subsec: reminder algebra}

In this paragraph, we consider a null region of spacetime $\mathcal{N}$ and $U \in \mathbb{R}$ an evolution parameter along it. We complement this coordinate by angles $(x^A)$. Let $\phi$ be a massless scalar field on $\mathcal{N}$. Its conjugate momentum is denoted $\pi$ and we have $\pi = \partial_U \phi$. As a genuine feature of null hypersurfaces, the momenta and the field are not independent variables.

\subsubsection*{Algebra of observables and GNS construction}

Consider the set of test functions on $\mathcal{N}$ namely the set of smooth functions of compact support on $\mathcal{N}$. This set is a vector space isomorphic to the space of solutions of the Klein-Gordon equation on a null hypersurface. Indeed, we assume that this equation reduces on $\mathcal{N}$ to $\partial_U \partial_V \phi = 0$ so\footnote{This is the case on the black hole Killing horizon and at null infinity.} that the general solution restricted to $\mathcal{N}$ is of the form $\phi = \phi(U, x^A)$. Let $\mathscr{S}$ be this space of solutions. From $f \in \mathcal{S}$ define the operator
\begin{equation}
    \label{eq: smearings of pi}
    \pi(f) = \int_{\mathcal{N}} \pi f \, \text{d}U \wedge \epsilon_S \, ,
\end{equation}
where $\epsilon_S$ is the volume form of the spatial cross-sections $S$ of $\mathcal{N}$. The operators \eqref{eq: smearings of pi} are generically unbounded so one considers instead their exponentiation. Introducing $W(f) := e^{i\pi(f)}$ we construct the \emph{Weyl algebra} of the theory
\begin{equation}
    \label{eq: weyl relations draft 2}
    W(f)W(g) = e^{\frac{i}{8}\Omega(f,g)}W(f+g) \quad \text{and} \quad W(f)^{\ast} = W(-f) \, ,
\end{equation}
where $\Omega$ is the symplectic form 
\begin{equation}
    \label{symplecticformsu}
    \forall (f, g) \in \mathscr{S}, \qquad \Om (f, g) = \int_{\mathcal{N}} (f \p_U g - g \p_U f) \text{d}U \wedge \eps_S \, .
\end{equation}
The set of $W(f)$ is a $\mathbb{C}^\ast$-algebra dubbed the \emph{algebra of observables} $\mathcal{A}$ of the spacetime region $\mathcal{N}$. The appearance of the symplectic structure in \eqref{eq: weyl relations draft 2} renders explicit the relation between the Weyl algebra and the equation of motion of the theory at hand. This structure is generically degenerate due to the zero modes in $U$ the test function may contain. To avoid this issue it is preferable to smear the momenta $\pi$ instead of the field $\phi$. The price to pay it that one cannot treat the soft sector of the theory at hand, a sector however irrelevant for the study we present in this work. Note that the algebra of observables corresponds precisely to the spacetime region in which the compact support of the test functions lies. Changing the latter changes the algebra of observables. An important example is the restriction of an algebra to the region above a cut $U = U_i$. From an algebraic point of view one gets a $\mathbb{C}^\ast$-subalgebra $\mathcal{A}_i \subset \mathcal{A}$ while from a spacetime perspective this means that the elements of the restricted algebra are operators $W(f)$ with $f$ of compact support in the region $U > U_i$.

The algebra of observables being a unital $\mathbb{C}^\ast$-algebra,\footnote{Meaning that the identity operator $1_\mathcal{A}$ belongs to it.} it admits Hilbert-spaces representations thanks to the Gelfand-Naimark-Segal (or GNS) theorem \cite{gelfand1943imbedding, segal1947irreducible}. In a nutshell one starts from an algebraic state i.e. a map $\omega: \mathcal{A} \to \mathbb{C}$ which is positive ($\omega(AA^\ast) \geq 0$ for all $A \in \mathcal{A}$) and normalized ($\omega(1_\mathcal{A}) = 1$), and get a $\omega$-dependent quadruple $(\mathscr{H}, \pi, \ket{\Omega}, \mathscr{D})_{\omega}$ where $\mathscr{H}_\om$ is a Hilbert space, $\pi_\om: \mathcal{A} \to \mathcal{L}(\mathscr{H}_\om)$ is a representation of the abstract algebra $\mathcal{A}$ in terms of linear operators acting on $\mathscr{H}_\om$, and $\mathscr{D}_\om = \pi_\om(\mathcal{A}) \ket{\Omega_\omega}$ a dense subset of $\mathscr{H}$ generated by the action of the algebra on the \emph{vacuum state} $\ket{\Omega_\omega}$. In  the following, in order to lighten the writing, the notation for the algebra elements $A \in \mathcal{A}$ is also used for their representatives $\pi_\om(A) \in \mathcal{L}(\mathcal{H}_\om)$. Also, one has that $\omega(A) = \bra{\Omega_\omega}A\ket{\Omega_\omega}$ for any element $A \in \mathcal{A}$. The GNS construction crucially depends on the initial choice of algebraic vacuum state. Two different initial states $\omega$ lead generically to two non-unitarily related GNS Hilbert-spaces. Finally, as the action of the algebra $\mathcal{A}$ on the vacuum $\ket{\Omega_\omega}$ generates the whole Hilbert space, one says that this vector is \emph{cyclic}. Given the exponentiation \eqref{eq: weyl relations draft 2}, once a vacuum state is chosen, the Weyl algebra is represented by a subalgebra of the algebra $\mathcal{B}(\mathscr{H}_\omega)$ of bounded operators of the GNS Hilbert space.

\subsubsection*{Modular theory and relative entropy}

There exist a special type of subalgebra of $\mathcal{B}(\mathscr{H}_\omega)$, the ones that are equal to their double commutant $\mathcal{A}''$. The latter are called \emph{von Neumann algebras}. Consider $\mathcal{A}_i$ the algebra of observables of the region $U > U_i$. Its commutant $\mathcal{A}_i^{'}$ corresponds to the algebra of observables of the region $U < U_i$. Therefore the double commutant is again $\mathcal{A}_i$ which is therefore a von Neumann algebra. In more general situations, given an algebra of observables, one associates directly to it a von Neumann algebra by taking the double commutant. Von Neumann algebras can be of three different types depending on whether or not one has pure states and density matrices (type I), only density matrices (type II) or neither of both (type III), as concepts such as traces are ill-defined \cite{witten2018aps, Sorce:2023fdx}. For example consider the vacuum state $\ket{\Omega_U}$ defined on the Weyl algebra associated to a non expanding null hypersurface $\mathcal{N}$ charted by the null coordinate $U$, that is defined so that $\ket{\Omega^U}$ is a Gaussian state on the subspace of positive frequency solutions induced by $U$. Via the GNS construction with this vacuum state, the algebra of observables $\mathcal{A}$ is a subalgebra of $\mathcal{B}(\mathscr{H}_{\Omega^U})$. As $\ket{\Omega^U}$ is a pure state wrt $\mathcal{A}$, $\,\mathcal{A}''$ ($= \mathcal{A}$) is a type I von Neumann algebra. Instead $\mathcal{A}_i^{''}$ is type III as $\ket{\Omega^U}$ becomes a mixed state on this subalgebra, and even looks like a thermal state with respect to some modular Hamiltonian
(see Appendix A of \cite{RBVilatte251}).

Up to taking the double commutant, we now consider that the algebra of observables $\mathcal{A}$ and any of its restrictions $\mathcal{A}_i$ are von Neumann algebras. To define the relative entropy we need to consider states $\ket{\Psi}$ which are cyclic (see the definition above) but also separating. A \emph{separating state} is a state such that for any $A\in \mathcal{A}$, $A\ket{\Psi} = 0$ implies $A = 0$. Physically, a state which is cyclic and separating is a state from which not only all other states in the Hilbert space can be generated (up to a limiting procedure) by applications of observables on it, but it is also a state which allows one to distinguish among all the observables in the algebra. Such states are at the heart of Tomita-Takesaki modular theory \cite{takesaki1972conditional, takesaki2003theory}. Given a cyclic and separating state $\ket{\Psi}$, one defines the \emph{modular operator} $\Delta_\Psi := S_\Psi^{\dagger} S_\Psi$ where $S_\Psi$ is the \emph{Tomita operator}
\begin{equation}
    \label{eq: Tomita operator draft 2}
    \forall A \in \mathcal{A} \, , \, S_{\Psi}\left(A \ket{\Psi}\right) = A^\dagger \ket{\Psi} \, ,
\end{equation}
and $S^\dagger_\Psi$ its Hermitian conjugate. One can show that the latter is actually the Tomita operator for the commutant algebra $\mathcal{A}'$. When traces and density matrices are defined, assuming that $\ket{\Psi}$ is cyclic and separating for $\mathcal{A}$, we consider $\hat{\rho}_{\Psi}^{\mathcal{A}_I}$ the density matrix associated to the restriction of $\ket{\Psi}$ to a subalgebra $\mathcal{A}_I$. In that case the modular operator reads $\Delta_\Psi = \hat{\rho}_{\Psi}^{\mathcal{A}_I} \otimes  \left(\hat\rho_{\Psi}^{\mathcal{A}_{II}}\right)^{-1}$ with $\mathcal{A}_{II} := \mathcal{A}_I^{-1}$. Given two cyclic and separating states $\ket{\Psi}$ and $\ket{\Omega}$ (the latter being often chosen to be the vacuum state of a GNS Hilbert space $\mathscr{H}_\Omega$) one defines the \emph{relative modular operator} $\Delta_{\Psi|\Omega} := S^{\dagger}_{\Psi|\Omega} S_{\Psi|\Omega}$ where $S_{\Psi|\Omega}$ is the \emph{relative Tomita operator}
\begin{equation}
    \label{eq: relative Tomita draft 2}
    \forall A \in \mathcal{A}, \; S_{\Psi|\Omega} \left( A\ket{\Psi} \right) = A^{\dagger}\ket{\Omega} \, .
\end{equation}
When we can define density matrices $\hat{\rho}_{\Omega}^{\mathcal{A}_{I}}$ and $\hat{\rho}_\Psi^{\mathcal{A}_{II}}$, we find that $\Delta_{\Psi|\Omega} = \hat{\rho}_{\Omega}^{\mathcal{A}_I} \otimes \left(\hat{\rho}_{\Psi}^{\mathcal{A}_{II}}\right)$.

Due to the cyclic and separating character of $\ket{\Psi}$, the modular operator is positive-definite, therefore one can take its logarithm and define the \emph{modular Hamiltonian}
\begin{equation}
    \label{eq: def mod hamiltonian draft 2}
    \Delta_\Psi = e^{-K_{\Psi}} \Longleftrightarrow K_{\Psi} = -\ln \Delta_\Psi \, ,
\end{equation}
which usually does not belong to $\mathcal{B}(\mathscr{H}_{\Omega})$. When the total algebra of observables $\mathcal{A}$ is restricted to a subalgebra $\mathcal{A}_i$ whose commutant $\mathcal{A}_i^{'}$ is just the algebra of the region $U < U_i$, we can decompose the modular Hamiltonian into two pieces 
\be
\label{eq: def one sided mod ham}
    K_\Psi =  K_\Psi^{\mathcal{A}_i} - K_\Psi^{\mathcal{A}_{i}^{'}}
\ee
called the \emph{one-sided modular Hamiltonians}. The right-sided piece $K_\Psi^{\mathcal{A}_i}$ commutes with the algebra $\mathcal{A}_i^{'}$ while the left sided $K_\Psi^{\mathcal{A}_{i}^{'}}$ commutes with $\mathcal{A}_i$. Given $T_{UU}$ the doubly null component of the stress-energy tensor of the scalar field theory we quantize on $\mathcal{N}$, the one-sided modular Hamiltonian of the vacuum state $\ket{\Omega_U} \in \mathscr{H}_{\Omega_U}$ wrt the restricted algebra $\mathcal{A}_i$ is\footnote{A vacuum state is always cyclic for its Hilbert space, as a consequence of the GNS theorem. It may become separating when restricted to a subalgebra of observables.} (see Appendix E of \cite{RBVilatte251})
\begin{equation}
    \label{eq: normal ordered and stress tensor draft 2}
    K^{\mathcal{A}_i}_{\Omega_U} = 2\pi \int_{U_i}^{+\infty} \int_S (U-U_i) :T_{UU}:_{\Omega^U} \text{d}U \wedge \epsilon_S \, ,
\end{equation}
where the integral runs over the spacetime region associated to the algebra $\mathcal{A}_i$.\footnote{Therefore the total modular Hamiltonian wrt the full algebra would have been
\begin{equation}
    \label{eq: total normal ordered and stress tensor}
    K^{\mathcal{A}}_{\Omega_U} = 2\pi \int_{-\infty}^{+\infty} \int_S (U - U_i) :T_{UU}:_{\Omega^U} \text{d}U \wedge \eps_S \, .
\end{equation}}
Note that in \eqref{eq: normal ordered and stress tensor draft 2} appears the normal-ordered version of the stress-tensor, wrt to the vacuum state $\ket{\Om_U}$. When proving the dual GSL (or the plain GSL), one needs to relate the modular Hamiltonian to geometric quantities such as the black hole area or the Bondi mass. Care as therefore to be taken because, as we shall recall in a subsequent paragraph, these geometric quantities are related to integrals of the covariant stress-energy tensor, not its normal-ordered version. Additional terms will therefore appear. 

We recall the fundamental result that the cyclic and separating vector $\ket{\Psi}$ satisfies the \emph{Kubo-Martin-Schwinger (KMS) conditions} wrt $\mathcal{A}$ i.e. 
\begin{equation}
    \label{eq: KMS condition draft 2}
    \forall (A,B) \in \mathcal{A}^{2}, \; \bra{\Psi}\Delta_{\Psi}^{-1}A\Delta_{\Psi}B\ket{\Psi} := \bra{\Psi}e^{K_{\Psi}}A e^{-K_{\Psi}}B\ket{\Psi} = \bra{\Psi}BA\ket{\Psi} \, .
\end{equation}
Finally we quote Tomita's theorem which asserts that the modular flow preserves the von Neumann algebra $\mathcal{A}$ i.e. $\forall t \in \mathbb{R}, \; \Delta_{\Psi}^{it}\mathcal{A}\Delta_{\Psi}^{-it} \subset \mathcal{A}$. Hence we understand why a geometric modular flow must be generated by a boost Hamiltonian like \eqref{eq: normal ordered and stress tensor draft 2}, since the flow must vanish when $U = U_i$ in order to preserve the algebra $\mathcal{A}_i$ attached to the subregion $U > U_i$.

\subsubsection*{Araki's relative entropy}

Given two cyclic and separating states $\ket{\Psi}$ and $\ket{\Omega}$ (often taken to be the vacuum state) for a von Neumann algebra $\mathcal{A}$, we write Araki's definition \cite{araki1975relative} of the \emph{relative entropy} as
\begin{equation}
    \label{eq: relative entropy draft 2}
    S(\Psi|| \Omega) := \bra{\Psi}\left(-\ln \Delta_{\Psi|\Omega} \right) \ket{\Psi} \, .
\end{equation}
This quantity is positive and vanishes if $\ket{\Psi} = A\ket{\Omega}$ with $A \in \mathcal{A}'$ a unitary element. The relative entropy evaluates the difference between two quantum states given that we only posses the observables belonging to $\mathcal{A}$ to perform measurement in order to distinguish them. Given this interpretation one understands that if we restrict to a subalgebra of observables, making a difference between two states will become harder. This is the \emph{monotonicity of the relative entropy} between two von Neumann algebras $\mathcal{B} \subset \mathcal{A}$
\begin{equation}
        \label{eq: monotonicity draft 2}
        S_{\mathcal{A}}(\Psi || \Omega) \geq S_{\mathcal{B}}(\Psi||\Omega) \, .
    \end{equation}
All the proofs of the GSL or its dual version are rooted in this inequality since Wall's seminal work \cite{Wall:2011hj}. Finally it can be shown (see Appendix D of \cite{RBVilatte251} for the proof) that the relative entropy is related to the one-sided modular Hamiltonian of $\mathcal{A}_I$ via the formula\footnote{We present the case where traces exist, the formula is also valid for type III algebras (such as $\mathcal{A}_i^\scri$ at null infinity) as long as one \emph{defines} the renormalized von Neumann entropy as
\begin{equation} 
\label{renovonentqft draft 2}
    S_{\Psi \lvert \Om}^{\text{v.N.}, \mathcal{A}_i^\scri} := S(\Psi \lvert \lvert \Om) - \langle K_\Om^{\mathcal A_i^\scri} \rangle_\Psi \, .
\end{equation}
which is possible as long as $\ket{\Psi}$ is in the domain of the one-sided modular Hamiltonian $ K_\Om^{\mathcal A_i^{\scri}}$.}
\begin{equation}
    \label{eq: relative in terms of von neumann}
    S\left(\hat \rho_{\Psi}^{\mathcal{A}_I} \Big| \Big| \hat \rho_{\Omega}^{\mathcal{A}_I} \right)  = -S^{\text{v.N}, \mathcal{A}_I}_{\Psi|\Omega} + \langle K^{\mathcal{A}_I}_{\Omega} \rangle_{\Psi}
\end{equation}
where we see appearing the \emph{renormalized von Neumann entropy} of $\ket{\Psi}$ with respect to $\ket{\Omega}$
\begin{equation}
    \label{eq: def von neumann entropy}
    S^{\text{v.N}, \mathcal{A}_I}_{\Psi|\Omega} = \Tr \left(\rho_{\Omega}^{\mathcal{A}_I} \ln \rho_{\Omega}^{\mathcal{A}_I} \right) -  \Tr \left(\rho_{\Psi}^{\mathcal{A}_I} \ln \rho_{\Psi}^{\mathcal{A}_I} \right) \, .
\end{equation}

\subsubsection*{The spacetime approach}

The objective of the last three paragraphs was to present the basic mathematical notions underpinning the proof of the dual GSL. These were abstract considerations so we try to be more concrete now. Going back on the null hypersurface $\mathcal{N}$ we consider the space of solutions of the Klein-Gordon equation which we complexity into $\mathscr{S}^\mathbb{C}$ together with the symplectic structure which becomes the \emph{Klein-Gordon product}
\begin{equation}
        \label{KGproductdraft2}
    \forall (f, g) \in \mathscr{S}^{\mathbb{C}}, \qquad \Om^{\mathbb{C}} (f, g) = i\int_{\mathcal{N}} (\bar{f} \p_U g - g \p_U \bar{f}) \text{d}U \wedge \eps_S \, .
    \end{equation}
It is generically not an inner product, unless we select a projector $K: \mathscr{S}^\mathbb{C} \to \mathscr{S}^{>0}_K$ which divides $\mathscr{S}^\mathbb{C}$ into a direct sum between the space of \emph{positive frequency solutions} $\mathscr{S}^{>0}_K$ and the space of negative frequency solutions. On $\mathscr{S}^{>0}_K$, $\Omega^\mathbb{C}$ becomes an inner product. One can then take the Cauchy completion of $\mathscr{S}^{>0}_K$ and get the \emph{one-particle Hilbert space} $\mathscr{H}^K_1$ from which one gets the Fock space via the usual formula
\begin{equation}
    \label{eq: sym fock space draft 2}
    \mathscr{H}^K := \mathscr{F}_{s}(\mathscr{H}^{K}_1) = \bigoplus_{n = 0}^{\infty} \left(\text{sym} \bigotimes_{k=0}^{n} \mathscr{H}^{1}_K \right) \,,
\end{equation}
where the symmetrization is taken as we deal with bosonic fields. The vacuum state $\ket{\Om^U}$ is defined as the state that is annihilated by the set of operators
\be \label{annihilationop section 3}
    a(Kf) = \Om^{\mathbb{C}}(Kf, \phi) = \langle f, \phi \rangle_K
\ee
and is unique. The canonical commutation relations read
\begin{equation}
    \label{eq: cano comu relat with symplec form}
    \left[a(Kf) , a^{\dagger}(Kg)\right] = \Omega^{\mathbb{C}}\left(a(Kf), a^{\dagger}(Kg) \right) = \langle f, g \rangle_K \, .
\end{equation}
The construction is therefore entirely based on the choice of projector $K$ i.e. on the choice of positive frequency modes. For example, given that $U$ is a natural time parameter on $\mathcal{N}$ one can project on the normalized modes $\frac{Y^l_m(x^A) e^{-i\Omega U}}{\sqrt{4\pi \Omega}}$ with $\Omega > 0$. This means that the field is decomposed as such\footnote{We consider a basis of complex spherical harmonics such that $\bar Y^l_m = (-1)^m Y^l_{-m}$.}
\begin{equation}
    \label{eq: 4D decomposition}
   \hat \phi ( U, x^A) = \sum_{l = 0}^{+ \infty} \sum_{m= - l}^{m=+l} \int_{0}^{+ \infty} \frac{d \Omega}{\sqrt{4\pi \Omega}} \left(Y_m^l(x^A) \hat a_{\Omega l m} e^{-i \Omega U} + \bar{Y}_m^l(x^A) \hat a_{\Omega l m}^{\dagger} e^{i \Omega U}\right) \, ,
\end{equation}
so that $\ket{\Omega^U}$ is the state annihilated by all $\hat{a}_{\Omega l m}$ with $\Omega > 0$. Of course other choices are possible therefore other vacua can be considered (see subsection \ref{subsec: HH and regularizations}).

The link with the more algebraic approach is done as follows. The Hilbert space $\mathscr{H}^K$ is actually the GNS Hilbert space built upon the algebraic quasifree state $\omega_U$ represented by the vector $\ket{\Omega^U}$. The vacuum state $\om_U$ is chosen so that it matches some boundary conditions at late time. In all the cases studied in \cite{RBVilatte251}, the choice of algebraic quasifree state $\omega$ from which one starts a GNS construction gets translated into a choice of positive frequency modes i.e. a choice of notion of time. However some naive choices of vacua may turn out to be irrelevant, as they lead to an inaccurate description of the physical process at hand, while more involved ones are more appropriate.

This concludes this subsection whose aim was to gather at one place all the notions the article needed to become (up to the proof of all the above statements) self-contained.

\subsection{A massless scalar field on maximally extended null infinity}
\label{subsec: reminder max scri}

\subsubsection*{Defining maximally extended null infinity}

\begin{figure}[ht]
    \centering
    \includegraphics[width=0.7\linewidth]{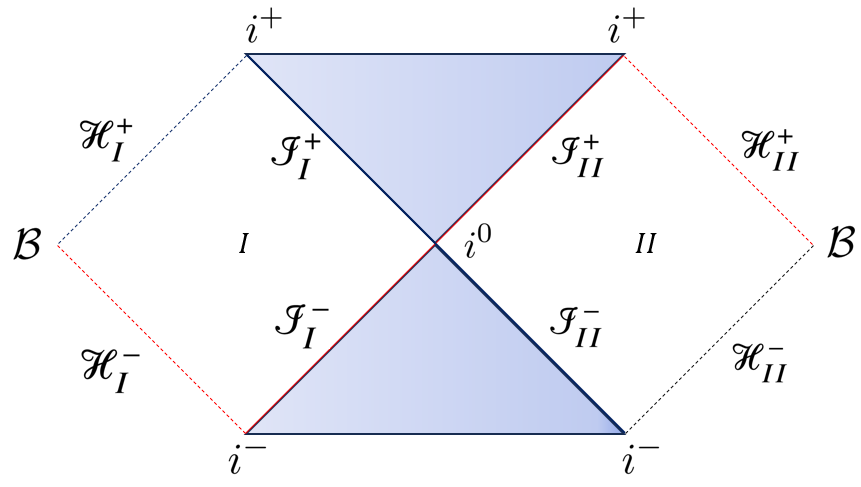}
    \caption{Conformal extension of the black hole spacetime centered on spacelike infinity. The blue region is the region $\mathcal{R}_\infty$ of \cite{Faulkner:2024gst}. Figure taken from the first part of this work \cite{RBVilatte251}.}
    \label{fig: gluing at iota zero}
\end{figure}

Consider the maximal extension of the Schwarzschild solution in $D = 4$ dimensions, charted by the Kruskal coordinates $(\tilde U,\tilde V, x^A)$ with $A = 1,2$. The associated Penrose diagram is centered on the bifurcation surface $\mathcal{B}$ located at $\tilde U = \tilde V = 0$. The coordinate $\tilde U$ (resp. $\tilde V$) is affine and inertial on the left (resp. right) horizon $\mathcal{H}_L = \mathcal{H}_{II}^- \cup \mathcal{H}_{I}^+$ (resp $\mathcal{H}_R = \mathcal{H}_{I}^- \cup \mathcal{H}_{II}^+$). Among all the isometries of this spacetime, the relevant one for our study is the one which is timelike in the external regions and the one wrt which the future event horizon is a Killing horizon. Its expression in terms of Kruskal coordinates is
\begin{equation}
\label{timelikekilling draft2}
    \xi = \kappa (\tilde{V} \partial_{\tilde{V}} - \tilde{U} \partial_{\tilde{U}}) \, ,
\end{equation}
with $\kappa = \frac{1}{4M}$ the black hole surface gravity. This setup is adapted to the GSL, which focuses on the dynamics on the future event horizon. The dual GSL instead is related to the dynamics of $\scri^+_I$. To translate Wall's proof there, we need to define a reference state (the vacuum state of the previous paragraph) which becomes thermal once restricted to the subalgebra of observables of $\scri^+_I$. To simplify the discussion by considering future directed time variables on $\scri^+_I$, we proposed in \cite{RBVilatte251} a slight change of perspective by considering the inverse Kruskal coordinates $U$ and $V$ defined by 
\begin{align}
\label{eq: def of U}
    U &:= -\frac{1}{\tilde{U}} = \left\{ 
    \begin{array}{ll}
        e^{\kappa u_+} & \mbox{on} \quad \scri^+_I \\
        - e^{-\kappa u_-} & \mbox{on} \quad \scri^-_{II}
    \end{array}
\right. \\ 
\label{eq: def of V}
V &:= -\frac{1}{\tilde{V}} = \left\{ 
    \begin{array}{ll}
        -e^{-\kappa v_+} & \mbox{on} \quad \scri^-_I \\
        e^{\kappa v_-} & \mbox{on} \quad \scri^+_{II}
    \end{array}
\right. \, ,
\end{align}
where $u_\pm$ (resp. $v_\pm$) are affine coordinates of $\scri^+_I$ and $\scri^-_{II}$ (resp. $\scri^-_I$ and $\scri^+_{II}$). The region covered by $U$, namely $\scri_R = \scri^+_I \cup \scri^-_{II}$ is called \emph{right maximally extended null infinity} while $V$ covers $\scri_L = \scri^-_I \cup \scri^+_{II}$ called \emph{left maximally extended null infinity} $\scri_L$. The set $(U, V, x^A)$ covers a conformal extension of the black hole background and one can draw a Penrose diagram which will now be centered on spacelike infinity $\iota^0$ located at $U = V = 0$ while the bifurcation surface is moved at $U = V = \infty$, see Figure \ref{fig: gluing at iota zero}.
The locus of interest for the remaining of the paper is $\scri_R$, that is, in all the considerations of the subsection \ref{subsec: reminder algebra}, we take $\mathcal{N} = \scri_R$ and consider on it a massless scalar field that we shall quantize in subsection \ref{subsec: HH and regularizations}. In particular we denote $\mathcal{A}^{\scri}$ the algebra of observables and by $\mathcal{A}_i^\scri$ the restricted algebra above a cut $U = U_i$. Note finally that the Killing field \eqref{timelikekilling draft2} now reads
\begin{equation}
\label{eq: Killin inverse kruskal}
    \xi = \kappa(U \partial_U - V \partial_V ) \, .
\end{equation}

\subsubsection*{Semiclassical Einstein dynamics}

As explained around \eqref{eq: normal ordered and stress tensor draft 2}, the integral of the stress-energy tensor is related to the variation of the relevant geometric quantities: $A$ the area of the horizon and $M$ the Bondi mass. The relation is obtained via the semiclassical dynamics on the perturbed horizon and at null infinity. 
\begin{itemize}
    \item \textbf{On the horizon} $\mathcal{H}_L$ the relevant equation is the Raychaudhuri equation \cite{Raychaudhuri}, which in the semiclassical regime reads
    \be \label{secondrayperphordraft2}
    \frac{d^2 \eps_S}{d\tilde V^2} = - 8 \pi \langle T_{\tilde V \tilde V} \rangle_\Psi \eps_S + O(\hbar^2)
    \ee
    where $\langle \cdot\rangle_\Psi$ denotes the expectation value in a state $\ket{\Psi}$ and $\epsilon_S$ is the volume form of a spatial cross-section of the horizon. Eq. \eqref{secondrayperphordraft2} has been obtained under the hypothesis of a small perturbation so that only linear terms remain. In particular 
    the expansion square $\theta^2$ is neglected in front of $\langle T_{\tilde V \tilde V}\rangle_\Psi$. One can show that at the end of the day we can write (here $\mathcal{D}_0^{\mathcal{H}} = (\tilde V_0, + \infty) \times S^2$)
    \be \label{differenceareasde1draft2}
   \f14 (A\rvert_{+ \infty} - A \rvert_{\tilde V = \tilde V_0}) = 2 \pi \int_{\mathcal{D}_0^\mathcal{H}} (\tilde V - \tilde V_0) \langle T_{\tilde V \tilde V} \rangle_\Psi \text{d}V \wedge \eps_S := 2 \pi \langle K_{\tilde V_0} \rangle_\Psi \, ,
    \ee
    as long as $\underset{\tilde V \rightarrow + \infty}{\lim} \tilde V \theta_l  = 0$. We refer to the quantity $K_{\tilde V_0}$ as the \emph{boost energy} above the cut $V_0$. From \eqref{differenceareasde1draft2} we get that between two cuts $\tilde V = \tilde V_1$ and $\tilde V = \tilde V_2 > \tilde V_1$ 
    \be
        \f14(A_{\tilde V_2} - A_{\tilde V_1}) = -2 \pi \left(\langle K_{\tilde V_2} \rangle_\Psi - \langle K_{\tilde V_1} \rangle_\Psi \right) \, .
    \ee

    \item \textbf{At null infinity} $\scri_R^+$ the relevant dynamics lies into Bondi's equation \cite{Bondi:1960jsa, sachs1962asymptotic} for the mass i.e. 
    \be \label{bondimasslossform2draft2}
    \frac{d M}{d u} = - \int_S \langle T_{uu} \rangle_\Psi \, \eps_S.
    \ee
    where we already considered the semiclassical regime. The equation is naturally defined on $\scri_R^+$ only, hence the use of the affine coordinate $u$ (corresponding to $u_+$ in the last paragraph). Writing \eqref{bondimasslossform2draft2} with our $U$ coordinate we get
    \be \label{Bml33hscridraft2}
        \kappa U^2 \frac{d^2 M}{dU^2} = \int_S \left(-\kappa^{-1} \p_u \langle T_{uu} \rangle_\Psi +  \langle T_{uu} \rangle_\Psi \right) \, \eps_S
    \ee
    and we are naturally led to assume that 
    \be \label{assumpthordraft2first}
        \kappa^{-1} \p_u \langle T_{uu} \rangle_\Psi \ll \langle T_{uu} \rangle_\Psi,
    \ee
    which is the equivalent on a black hole background of the condition $\theta^2 \ll \langle T_{\tilde V \tilde V} \rangle_\Psi$. The condition \eqref{assumpthordraft2first} is fundamental for proving the dual GSL and it possesses an interesting thermodynamic interpretation, see subsection \ref{subsec: time scales}. At the end of the day the link between the Bondi mass and the boost energy at $\scri_R$ is given by
    \be \label{kuzerodefdraft2}
     M \lvert_{+ \infty} - M (U_0) = - \int_{\mathcal{D}^{\scri}_0} \kappa (U - U_0) \langle T_{UU} \rangle_\Psi \text{d}U \wedge \eps_S := - \kappa \langle K_{U_0} \rangle_\Psi
    \ee
    where $\mathcal{D}^{\scri}_0 = (U_0, +\infty) \times S^2$, so that between to finite cuts $U = U_1$ and $U= U_2 > U_1$ one has 
    \be \label{diffbondimassdraft2}
        M (U_2) - M(U_1) = \kappa (\langle K_{U_2} \rangle_\Psi - \langle K_{U_1} \rangle_\Psi) \, .
    \ee
    Note the analogy between the dynamics of the Killing horizon $\mathcal{H}_L^+$ and the one of $\scri_R^+$. One can push it further by considering the boost field $\chi = \kappa(U-U_0)\partial_U = \left(1 - e^{-\kappa(u - u_0)}\right)\partial_u$. At the cut $U = U_0$ it vanishes while at late times $U \to +\infty$ it reduces to the affine Killing time $\partial_u := \kappa U\partial_U$. Note finally that the adapted time $\bar u$ for quantization above a cut $U = U_0$ is\footnote{An adapted time for quantization in a null hypersurface $\mathcal{N}$ is a variable that runs from $-\infty$ to $+\infty$ as one goes along $\mathcal{N}$.}
    \be \label{modaffinetimeu0draft2}
        \bar{u} := \kappa^{-1} \ln{(U - U_0)} = u + \kappa^{-1} \ln{(1 - e^{- \kappa(u - u_0)})} \, .
    \ee
    This consideration is crucial when dealing with the dual GSL in the Unruh state, see subsection \ref{subsec: construction Unruh}.
\end{itemize}


\subsection{The Hartle-Hawking state and its regularizations}
\label{subsec: HH and regularizations}

On $\scri_R$, whose evolution parameter is the inverse Kruskal time $U$, the natural decomposition of the field $\phi$ is the one of \eqref{eq: 4D decomposition}. However, having at hand a bunch of different angular-momentum modes $(l,m)$ allows for more freedom. In particular we motivated in \cite{RBVilatte251} a more general decomposition of the scalar field operator of the form 
\begin{equation}
    \label{eq: full quantum fieldlm draft 2}
   \hat \phi ( U, x^A) = \sum_{l = 0}^{+ \infty} \sum_{m= - l}^{m=+l} \int_{0}^{+ \infty} \frac{d \Omega}{\sqrt{4\pi \Omega}} \left(Y_m^l(x^A) \hat a_{\Omega l m} e^{-i \Omega U^{(l,m)}(U)} + \bar{Y}_m^l(x^A) \hat a_{\Omega l m}^{\dagger} e^{i \Omega U^{(l,m)}(U)}\right) \, ,
\end{equation}
with $U^{(l,m)}$ a notion of time attached to the specific angular mode $(l,m)$ one considers. In particular it can be different between two modes i.e. $U^{(l,m)} \neq U^{(l',m')}$ if $(l,m) \neq (l',m')$. The modes $Y_m^l \frac{e^{- i \Om U^{(l,m)}}}{\sqrt{4 \pi \Om}}$ are normalized and define the positive frequency solutions (when $\Omega > 0$). The vacuum state $\ket{\Omega^{\{U^{(l,m)}\}}}$ is therefore the state annihilated by all operators $\hat{a}_{\Omega l m}$ i.e.
\be
\label{lmvacdraft2}
    \forall \Om > 0, \quad \forall (l,m), \qquad \hat a_{\Om l m} \ket{\Omega^{\{U^{(l,m)}\}}} = 0 \, .
\ee
Finally the one-particle Hilbert space is obtained through the repeated action of the creation operators $\hat{a}^{\dagger}_{\Omega l m}$ on \eqref{lmvacdraft2}, and the Fock space via \eqref{eq: sym fock space draft 2}. The vacuum state is invariant under M\"obius reparameterization of the times $U^{(l,m)}$ i.e.
\be \label{mobius4Ddraft2}
    \forall (l,m):
    U^{(l,m)} \rightarrow \tilde{U}^{(l,m)} = \frac{a_{lm} U^{(l,m)} + b_{lm}}{c_{lm} U^{(l,m)} + d_{lm}}, \qquad a_{lm} d_{lm} - b_{lm} c_{lm} \neq 0,
\ee
which can be shown either using directly the field decomposition \eqref{eq: 4D decomposition} or by analyzing the transformation properties of the two point function (see \cite{RBVilatte251}).

As the notion of time depends on the angular momentum mode, it is expected that the total vacuum state $\ket{\Omega^{(l,m)}}$ admits a tensorial decomposition wrt to the modes $(l,m)$. Indeed if one considers
\be \label{philmdecompdraft2}
    \hat \phi = \sum_{lm} \hat \phi_{lm}, \qquad \hat \phi_{lm} = \int_{0}^{+ \infty} \frac{\text{d} \Om}{\sqrt{4 \pi \Om}} \left(Y_m^l \hat a_{\Om l m} e^{- i \Om U^{(l,m)}} + \bar{Y}_m^l \hat a_{\Om l m}^\dag e^{i \Om U^{(l,m)}}\right) \, ,
\ee
the $\hat{\phi}_{lm}$ become effectively two-dimensional chiral fields. Given the commutation relations
\be
\label{eq: chiral commut relations}
    \left[\phi_{lm}(U), \pi_{l'm'}(U')\right] = \frac{i}{2} \delta(U - U') \delta_{ll'} \delta_{mm'} \bar{Y}_{m'}^{l'} Y_{m}^l \, ,
\ee
the two-dimensional chiral CFT attached to two different modes $(l,m) \neq (l',m')$ are independent. Calling $\ket{\Omega^{U^{(l,m)}}}$ the vacuum state for the theory $\hat{\phi}_{lm}$ we can write the total vacuum $\ket{\Omega^{\{U^{(l,m)}\}}}$ as 
\be \label{vallmtendraft 2}
    \ket{\Omega^{\{U^{(l,m)}\}}} = \bigotimes_{lm} \ket{\Omega^{U^{(l,m)}}}\, ,
\ee
so that $\{U^{(l,m)}\}$ can be seen as a sequence of times $U^{(l,m)}$, indexed by $l$ and $m$. When decomposing the field in $(l,m)$ modes like in \eqref{philmdecompdraft2} it can be useful to introduce also the notation $\hat{\phi}_l = \sum_m \hat{\phi}_{lm}$ so that the $l$-stress-energy tensor (renormalized wrt the vacuum $\ket{\Omega^{U^{(l,m)}}}$ reads
\begin{equation}
    \label{eq: lstresstensor}
    :T_{U^{(l,m)} U^{(l,m)}}^{l}:_{\Omega^{U^{(l,m)}}} = \partial_{U^{(l,m)}}\hat{\phi}_{lm} \partial_{U^{(l,m)}}\hat{\phi}_{lm} - \langle \partial_{U^{(l,m)}}\hat{\phi}_{lm} \partial_{U^{(l,m)}}\hat{\phi}_{lm} \rangle_{\Omega^{U^{(l,m)}}} \, .
\end{equation}

In \cite{RBVilatte251}, three classes of vacuum states of the form \eqref{lmvacdraft2} have been considered.
\begin{itemize}
    \item \textbf{The Hartle-Hawking state} at $\scri_R$, which is built on the times
    \begin{equation}
        \label{eq: HH time lm}
        \forall (l,m) \, , \, U^{(l,m)} = U \, ,
    \end{equation}
    and denoted $\ket{\Omega_H}$. Such a state is pure wrt the algebra $\mathcal{A}^{\scri}$ while it becomes cyclic and separating for the restricted algebra $\mathcal{A}_i^\scri$. Therefore, given the symmetries \eqref{mobius4Ddraft2}, it is also cyclic and separating for any restriction above any cut i.e. wrt $\mathcal{A}_i^\scri$.\footnote{To be more precise the state built upon \eqref{eq: HH time lm} coincides with the Hartle-Hawking state only on the subalgebra $\mathcal{A}_0$ (and other restrictions), but not on the total algebra $\mathcal{A}$.} It can be used to model a black hole at equilibrium with its radiation. At any point of the bulk spacetime, the net energy flux is vanishing, as the outgoing radiation is exactly compensated by incoming radiation from $\scri_L^-$. This is not true at $\scri_R^+$ where an infinite energy flux is observed. These undesirable divergences betray the impossibility to describe accurately a radiating Schwarzschild black hole using that state: an asymptotic observer won't actually be able to distinguish the black hole from the surrounding radiation. Moreover, it is unrealistic to think that all the angular modes $(l,m)$ can make it to future null infinity. Due to the potential barrier which depends on $l$,\footnote{Recall that the Schwarzschild potential takes the form
    \begin{equation}
        \label{eq: scharz potential}
        V_l(r) =  \left(1 - \frac{2M}{r}\right)\left(\frac{l(l+1)}{r^2} + \frac{2M}{r^{3}} \right) \, .
    \end{equation}} high-angular momentum modes should be back-scattered inside the black hole and therefore be inaccessible to an asymptotic observer. This is why in \cite{RBVilatte251} we proposed two regularizations of the Hartle-Hawking state.

    \item \textbf{The $L$-vacuum} corresponds to a \emph{hard regularization} in which a cutoff $L < +\infty$ is chosen a priori. We denote it by $\ket{\Omega_H^L}$. The modes $l < L$ corresponds to thermal excitations at Hawking temperature $T_H$ at future null infinity while the modes $l \geq L$ have been completely back-scattered, therefore the observer sees the Minkowski vacuum for them. Physically, it could be realized by feeding the black hole with some incoming thermal radiation made of modes of angular momentum $l \leq L$ at the Hawking temperature. If $L$ is large enough, the ingoing radiation exactly compensates the outgoing radiation (because the modes $l \geq L$ emitted from the black hole cannot cross the potential barrier) and the field is in the $L$-vacuum. The associated notion of time is 
    \be \label{timehhbigULvacdraft2}
    U^{(l,m)} = \left\{ 
    \begin{array}{ll}
        U  & \mbox{if} \quad l < L \\
        u_{\pm} = \kappa^{-1}\ln{\pm U} & \mbox{if} \quad  l \geq L
    \end{array}
    \right.
    \ee
    while the space of positive frequency solutions is given by
    \begin{equation}
    \label{eq: L-vac positive solutions draft 2}
        \mathscr{S}^{>0}_{L} :=\left\{f \in \mathscr{S}^{\mathbb{C}} \Bigg| f(U, x^A) = \sum_{l=0}^{L-1} \sum_{m = -l}^{l} \int_{0}^{+\infty} \text{d}\Omega a_{\Omega lm} Y^{l}_{m}e^{-i\Omega U} + \sum_{l=L}^{+\infty} \sum_{m = -l}^{l} \int_{0}^{+\infty} \text{d}\omega a_{\omega lm} Y^{l}_{m}e^{-i\omega u_\pm} \right\} \, .
    \end{equation}
    Thanks to the cutoff $L$, only a finite number of modes are excited relative to the Minkowski vacuum, rendering therefore the energy density and the energy flux finite at future null infinity.\footnote{Recall that $L = 0$ corresponds to the Minkowski vacuum while the Hartle-Hawking state is recovered in the limit $L \to + \infty$.} 

    \item \textbf{The $\kappa_l$-vacuum} $\ket{\Omega^{ \{ \kappa_l \} }}$ corresponds to a \emph{soft regularization} in which we take into account that each angular mode $l$ is effectively scattered differently by the black hole. We translated this observation in \cite{RBVilatte251} by proposing to associate to the modes belonging to the same sector $l$ their own effective temperature
    \be \label{effectivetldraft2} 
        T_l := \frac{\kappa_l}{2 \pi}.
    \ee
    where $\kappa \geq \kappa_l \geq 0$ and $\kappa_l \to 0$ when $l \to + \infty$, as the bigger the angular mode, the more back-scattered. The sequence $\{ \kappa_l \}$ is not precised, but, given the shape of the potential \eqref{eq: scharz potential}, an interesting candidate to model an evaporating black hole might be $\kappa_l = e^{-\alpha l(l+1)}$ with $\alpha > 0$ a constant. The associated notion of time is 
    \be \label{ulmsoftdef}
        U^{(l,m)} = \left\{ 
        \begin{array}{ll}
            U^l := e^{\kappa_l u_+} & \mbox{on} \quad \scri^+_R \\
            U^l := -e^{-\kappa_l u_-} & \mbox{on} \quad \scri^-_R
        \end{array}
    \right.
    \ee
    so that $U^l = U$ if $\kappa_l = \kappa$ and $ \lvert U^l \lvert = \lvert U \lvert^{\frac{\kappa_l}{\kappa}}$, while the set of positive frequency solutions is given by
    \begin{equation}
        \label{eq: positive space kappa vacuum}
        \mathscr{S}^{>0}_{\kappa_l} := \left\{f \in \mathscr{S}^{\mathbb{C}} \Bigg| f(U,x^A) = \sum_{l=0}^{+\infty} \sum_{m=-l}^{l} \int_{0}^{+\infty} \text{d}\Omega \, a_{\Omega l m}Y^{l}_{m}e^{-i\Omega U^l}\right\} \, .
    \end{equation}
    To each sequence of $\{\kappa_l\}$ is associated a vacuum state $\ket{\Omega^{ \{ \kappa_l \} }}$ and changing at least one of the $\kappa_l$ lead to a different vacuum whose associated Hilbert space $\mathscr{H}_{\Omega^{\{\kappa_l\}}}$ is generically non unitarily related to the previous one. However, thanks to the freedom one has in choosing the effective temperature, we believe that this state may lead to a more accurate description of a radiating Schwarzschild black hole, since the gray-body factors depend on the angular momentum $l$. Therefore, the class of $\kappa_l$-vacua constitutes nice toy models to study some thermodynamic properties of the Unuruh vacuum. In particular the bunch of effective temperatures \eqref{effectivetldraft2} allows to extract work from 
    the $\kappa_l$-vacua, see Section \ref{sec: work extraction}.
\end{itemize}
This concludes our review of the relevant vacuum states considered in \cite{RBVilatte251}, to which we refer the interested reader for further details. We now turn to a recap of the proof of the dual generalized second law.

\subsection{Relative entropy and dual GSL}
\label{subsec: reminder dual GSL}

\begin{figure}[ht]
    \centering
    \includegraphics[width=0.5\linewidth]{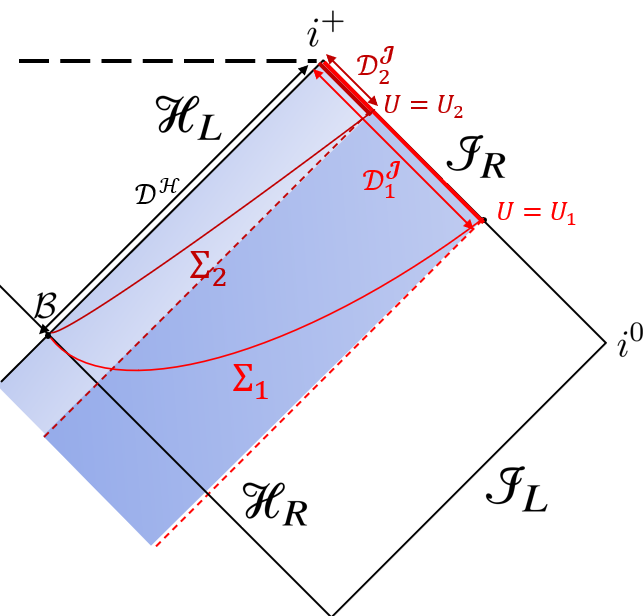}
    \caption{Setup in which the hypersurfaces $\Sigma_1$ and $\Sigma_2$ both starts at the horizon bifurcation surface $\mathcal{B}$ and end at different cuts $U = U_1$ and $U = U_2$ respectively at $\scri_R^+$. Are also depicted the regions $\mathcal{D}^{\mathcal{H}}$ and $\mathcal{D}_i^{\scri}$. Figure taken from \cite{RBVilatte251}.}
    \label{fig: semi setup GSL}
\end{figure}

We consider two spacelike hypersurfaces $\Sigma_1$ and $\Sigma_2$ starting both at the bifurcation surface $\mathcal{B}$ and ending at two different cuts at $\scri_R^+$, say $U = U_1 > 0$ and $U = U_2 > U_1$, $\mathcal{A}_2^{\scri} \subset \mathcal{A}_1^{\scri}$ . A representation of this situation is depicted in Figure \ref{fig: semi setup GSL}.
We push these hypersurfaces to infinity so that they cover a region $\mathcal{D}_i^\mathcal{B}$ of the union $\mathcal{H}_L^+ \cup \scri_R^+$. We have that
\begin{equation}
    \label{eq: def domain total draft 2}
    \mathcal{D}_i^{\mathcal{B}} = \underbrace{\left( (0, +\infty) \times S^2_{\mathcal{H}}\right)}_{:= \mathcal{D}^{\mathcal{H}}} \cup \underbrace{\left( (U_i, +\infty) \times S^2_{\scri}\right)}_{:= \mathcal{D}_{i}^{\scri}} \, ,
\end{equation}
and the dual GSL between $\mathcal{D}_1^{\mathcal{B}}$ and $\mathcal{D}_2^{\mathcal{B}}$ is equivalent to the one between $\Sigma_1$ and $\Sigma_2$ by unitarity. We denote $\mathcal{A}_i^{\scri}$ the algebra of observables of $\mathcal{D}_i^{\scri}$ and similarly $\mathcal{A}^{\mathcal{H}}$ is the one of $\mathcal{D}^{\mathcal{H}}$ so that the total algebra of $\mathcal{D}_i^{\mathcal{B}}$ is $\mathcal{A}_i^{\mathcal{B}} = \mathcal{A}^{\mathcal{H}} \otimes \mathcal{A}_i^{\scri}$.

Take a vacuum state $\ket{\Omega}$, which represents the asymptotic state to which our system (the radiating black hole) will relax at late times, and construct its GNS Hilbert space $\mathscr{H}_\Omega$. With respect to any restricted algebra $\mathcal{A}_i^\scri$ this state is assumed to be cyclic and separating. Consider $\ket{\Psi}$ a second cyclic and separating state for all $\mathcal{A}_i^\scri$ so that the monotonicity of the relative entropy \eqref{eq: monotonicity draft 2} reads
\begin{equation}
    \label{eq: consequence relat entropy draft 2}
    S_{\mathcal{A}_1^{\scri}}(\Psi || \Omega) \geq S_{\mathcal{A}_2^{\scri}}(\Psi||\Omega) \, .
\end{equation}
since $\mathcal{A}_2^{\scri} \subset \mathcal{A}_1^{\scri}$. Using \eqref{eq: relative in terms of von neumann} we get
\begin{equation}
    \label{eq: fundamental inequality draft 2}
    \Delta S^{\text{v.N.}, \mathcal{A}_i^{\scri}}_{\Psi|\Omega} - \Delta \langle K^{\mathcal{A}_i^{\scri}}_{\Omega}\rangle_{\Psi} \geq 0 \, ,
\end{equation}
with
\begin{equation}
    \label{eq: def des delta draft 2}
    \Delta S^{\text{v.N.}, \mathcal{A}_i^{\scri}}_{\Psi|\Omega} := S^{\text{v.N.}, \mathcal{A}_2^{\scri}}_{\Psi|\Omega} - S^{\text{v.N.}, \mathcal{A}_1^{\scri}}_{\Psi|\Omega} \quad \text{and} \quad \Delta \langle K^{\mathcal{A}_i^{\scri}}_{\Omega}\rangle_{\Psi} := \Delta \langle K^{\mathcal{A}_2^{\scri}}_{\Omega}\rangle_{\Psi} - \Delta \langle K^{\mathcal{A}_1^{\scri}}_{\Omega}\rangle_{\Psi} \, .
\end{equation}
For the total algebra $\mathcal{A}_i^{\mathcal{B}} = \mathcal{A}^{\mathcal{H}} \otimes \mathcal{A}_i^{\scri}$ we just consider as vacuum state the product state 
\begin{equation}
    \label{eq: total vac draft 2}
    \ket{\bar \Omega} := \ket{\Omega^\mathcal{H}} \otimes \ket{\Omega} \, ,
\end{equation}
which is cyclic and separating for all $\mathcal{A}_i^\mathcal{B}$, together with $\ket{\bar \Psi} \in \mathscr{H}_{\bar \Omega}$ also cyclic and separating, so that at the end we get the \emph{fundamental inequality}
\begin{equation}
    \label{eq: total fundamental inequality}
    \Delta S^{\text{v.N.}, \mathcal{A}^\mathcal{B}_i}_{\bar \Psi|\bar \Omega} - \Delta \langle K^{\mathcal{A}_i^{\scri}}_{ \Omega}\rangle_{\bar \Psi} \geq 0 \, ,
\end{equation}
between the spacetime domain $\mathcal{D}^{\mathcal{B}}_1$ and $\mathcal{D}^{\mathcal{B}}_2$. Then we use \eqref{eq: normal ordered and stress tensor draft 2} to relate the one-sided modular Hamiltonian to the normal-ordered stress-tensor. It remains, finally, to recover the Bondi mass using the relation between the normal-ordered stress tensor and the covariant stress tensor. The dual generalized second law can then be formulated as the statement that a thermodynamic potential $\mathcal{G}_\Om$ decreases between $\Sigma_1$ and $\Sigma_2$ i.e.
\begin{equation}
    \label{eq: dual GSL general}
    \Delta \mathcal{G}_\Om \leq 0 \, .
\end{equation}
Now we specialize these steps to the three vacua we defined in the subsection \ref{subsec: HH and regularizations}.
\begin{itemize}
    \item \textbf{In the Hartle-Hawking state} the one-sided modular Hamiltonian wrt to the algebra $\mathcal{A}_i^{\scri}$ reads 
    \be \label{modhamnormordidraft2}
        K_{\Om_H}^{\mathcal{A}_i^\scri} = 2 \pi \int_{\mathcal{D}_i^\scri} (U - U_i) :T_{UU}:_{\Om_H} \text{d}U \wedge \eps_S \, ,
    \ee
    and the relation between $:T_{UU}:_{\Omega_H}$ and $T_{UU}$ is of the form
    \begin{equation}
        \label{eq: relation cov and normal}
        T_{UU} = \, :T_{UU}:_{\Omega_H} + \, \text{Schwarzian term} \, ,
    \end{equation}
    so that using Einstein's equations \eqref{diffbondimassdraft2} we get (the state $\ket {\bar \Psi}$ has to be chosen such that \eqref{assumpthordraft2first} holds)
    \begin{equation}
        \label{eq: modular HH and bondi mass}
        \Delta \langle K_{\Om_H}^{\mathcal{A}_i^{\scri}} \rangle_{\bar \Psi} = \Delta M - \Delta M_{\Om_H}
    \end{equation}
    with $\Delta M_{\Omega_H}$ the variation of Bondi mass in the Hartle-Hawking state. Given that this state is exactly thermal, one can apply the Clausius relation ($\Delta S = \frac{Q}{T_H}$) for the infinite bath of radiation at Hawking temperature $T_H$. As no work can be extracted from it, one has $W = 0$ and we get $\Delta S = \frac{\Delta E}{T_H}$ which gives in the case at hand
    \be \label{zeropointenergyentropyreldraft2}
        \Delta S_{\Om_H} = \frac{\Delta M_{\Om_H}}{T_H} \, ,
    \ee
    meaning that the zero point energy in the Hartle-Hawking state actually corresponds to a variation of entropy. Adding \eqref{zeropointenergyentropyreldraft2} to \eqref{eq: modular HH and bondi mass} we find from the inequality \eqref{eq: total fundamental inequality}
    \begin{equation}
        \label{eq: dual GSL HH}
        \Delta M - T_H \Delta S \leq 0 \Longrightarrow \mathcal{G}_{\Omega_H} = M - T_H S := \mathcal{F}
    \end{equation}
    with $\Delta S = \Delta S_{\bar{\Psi} \lvert \bar{\Om}_H}^{\text{v.N.},\mathcal{A}_i^\mathcal{B}} +  \Delta S_{\Om_H}$. The relevant potential is therefore the free energy, betraying the relation with the thermodynamics of a system which exchanges energy with a bath at Hawking temperature. However, recall that expressions appearing in \eqref{zeropointenergyentropyreldraft2} are divergent, hence the need for a regularization. One can first use the hard regularized $L$-vacuum $\ket{\Omega^L_H}$. It is straightforward to check that if $L$ is sufficiently large, the exact same steps can be repeated using now the Hilbert space $\mathscr{H}_{\Om^L_H}$ built from the $L$-vacuum $\ket{\Om_H^L}$,\footnote{It means that the larger $L$ is, the less restrictive are the conditions on the state $\ket{\bar \Psi} \in \mathscr{H}_{\bar \Om^L_H}$ needed to recover \eqref{eq: dual GSL HH}. In particular, it works immediately for an arbitrary state in the Fock space built from the modes with angular momentum $l \leq L$. See \cite{RBVilatte251} for further details.} but now the quantities appearing in \eqref{zeropointenergyentropyreldraft2} are finite, as are the two members of the left hand side of the inequality in \eqref{eq: dual GSL HH}.

     \item \textbf{In the $\kappa_l$-vacuum} the one-sided modular Hamiltonian can be written as 
     \be \label{modhamsoftregudraft2}
        K^{\mathcal{A}_i^\scri}_{\Om^{\{ \kappa_l \}}_H} = 2 \pi \sum_{l} \int_{U_i^l}^{+ \infty} \int_S (U^l - U_i^l) : T_{U^l U^l}^l :_{\Om^{\{ \kappa_l \}}_H} \text{d}U^l \wedge \epsilon_S \, ,
    \ee
    with $T^l_{UU}$ defined in \eqref{eq: lstresstensor}. For the proof to work we needed to choose the sequence $\{ \kappa_l \}_{l \geq 0}$ so that $\sum_l (2l+1)\kappa_l < +\infty$. Under that hypothesis, together with \eqref{assumpthordraft2first}, we found in \cite{RBVilatte251} the following inequality between the cut $u = u_1$ and the cut $u = u_2$ of $\scri_R^+$ 
    \be \label{grandpotential2var}
        \frac{\Delta M}{T_H} - \sum_{l = 0}^{+ \infty} \sum_{m = -l }^{m = +l} \int_{0}^{+ \infty} \frac{\m_{\om l}}{T_H} \langle \Delta n_{\om l m} \rangle_{\bar \Psi} d\om - \Delta S \leq 0 \, ,
    \ee
    with $\ket{\bar \Psi} \in \ket{\bar \Om_H^{\{\kappa_l\}}}$. In this expression we recognize \emph{chemical potential terms} $\mu_{\omega l}$ whose relation to $\kappa_l$ is
    \begin{equation}
        \label{eq: relation mu kappal}
        \mu_{\omega l} = \omega \left( 1 - \frac{\kappa}{\kappa_l} \right)
    \end{equation}
    which are negative (at it should for bosonic fields) as $\kappa \geq \kappa_l$. Also $\Delta n_{\omega l m}$ is the spectral density operator which reads
    \be
    \label{eq: def nomegalm}
     \Delta n_{\om l m} := \Delta \bar{n}_{\om l m} - (u_2 - u_1) \langle N_{\om l m} \rangle_{\Om_H^{\{ \kappa_l \}}} 
    \ee
    with on the one hand $\langle N_{\om l m} \rangle_{\Om_H^{\{ \kappa_l \}}}$ the vev of the number operator for the modes $(\omega l m)$ in the $\kappa_l$-vacuum and on the other hand $\bar{n}_{\om l m}$ the excess density of modes compared to the vacuum
    \begin{equation}
        \label{eq: excess density}
        \bar{n}_{\om l m} = \underset{\Delta \om \rightarrow 0}{\lim} \sum_{\om, \om + \Delta \om } \frac{N_{\om l m} - \langle N_{\om l m} \rangle_{\Om_H^{\{ \kappa_l \}}}}{\Delta \om} \, .
    \end{equation}
    Therefore in the $\kappa_l$-vacuum the relevant thermodynamic potential is the \emph{grand-potential} which reads
    \begin{equation}
        \label{eq: grand potential kappa l vac}
    \mathcal{G}_{\Omega^{\{\kappa_l\}}} := M - \sum_{l = 0}^{+ \infty}\sum_{m = -l}^{+l} \int_{0}^{+\infty} \mu_{\omega l} \langle n_{\omega l m} \rangle_{\bar \Psi} \text{d}\omega - T_H S \, .
    \end{equation}
    The non-geometric terms related to the chemical potentials are of particular interest as we shall show in the next section that they are related to work terms. As a final remark, which will help to make contact latter with the discussion of the Unruh vacuum, let us note that the modular Hamiltonian \eqref{modhamsoftregudraft2} can also be written in terms of the excess density \eqref{eq: excess density} and the chemical potentials \eqref{eq: relation mu kappal}
    \begin{equation}
        \label{eq: modhamkappal link unruh}
        K_{\Om_H^{\{\kappa_l\}}}^{\mathcal{A}_i^\scri} = 2 \pi \int_{\mathcal{D}_i^\scri} (U - U_0) :T_{UU}:_{\Om_H^{\{\kappa_l\}}} \text{d}U \wedge \eps_S - \sum_{lm} \int_{0}^{+ \infty} \frac{\m_{\om l}}{T_H} \bar{n}_{\om l m}  \text{d}\om \, ,
    \end{equation}
    showing even more explicitly the decomposition into two terms: a geometric contribution related to the Bondi mass and a non-geometric one associated to the chemical potentials.
\end{itemize}
This concludes our recollection of the fundamental definitions and result of the first part of this work \cite{RBVilatte251}. These results complete and generalize the earlier results obtained by one of us in \cite{ARB24}, and place them on a precise and rigorous footing beyond the effective approach employed there. Now we delve into the relation with thermodynamics of open systems.


\section{Relations with dynamics of open systems}
\label{sec: open systems}

\subsection{Characteristic time scales}
\label{subsec: time scales}

The assumption \eqref{assumpthordraft2first} was essential to relate the Bondi mass to the boost energy so in fine to the one-sided modular Hamiltonian \eqref{modhamnormordidraft2}. We aim to justify it by making the parallel between black hole thermodynamics at null infinity and the theory of open quantum systems more explicit. We can write \eqref{assumpthordraft2first} as 
\be \label{assumpthor2draft2}
    \frac{\tau_B}{\tau_D} \ll 1
\ee
where $\tau_B = \kappa^{-1} = 4 M$ and $\tau_D = \frac{\langle T_{uu} \rangle_\Psi}{\p_u \langle T_{uu} \rangle_\Psi}$. The latter corresponds to the typical time evolution of the stress energy tensor in a state $\ket{\Psi}$. In the case where we have a constant flux of particles at $\scri^+_R$, $\tau_D = \infty$ and \eqref{assumpthor2draft2} is obviously satisfied. In fact, this assumption is really analogous to some assumptions on the relaxation time of the bath that we encounter in open quantum systems.
In this latter framework, one considers a system—corresponding, in the quantum field theory picture, to a quantum field in a given state $\ket{\Psi} $ on $\scri_R$—that relaxes toward an equilibrium state under the influence of a reservoir. The equilibrium state corresponds to a vacuum state, such as the Hartle–Hawking vacuum, the Unruh vacuum, or one of the regularized vacua introduced in subsection~\ref{subsec: HH and regularizations}. In the gravitational setting, the analogue of the reservoir is provided by the spacetime dynamics itself—for instance, a black hole collapse—which settles into specific late-time boundary conditions. In our approach, however, since we quantize directly on $\scri_R$, the choice of late-time boundary conditions becomes a kinematical choice rather than a dynamical one. Indeed, it is equivalent to selecting a vacuum state—and hence a Hilbert space—among the infinitely many unitarily inequivalent representations of the algebra of observables on $\scri_R$.\footnote{This multiplicity arises from the breakdown of the Stone–von Neumann theorem in quantum field theory.} Different choices of vacuum states therefore correspond to different late-time boundary conditions. Within the theory of open quantum systems, a particularly important class of systems is provided by Markovian systems, namely systems whose evolution is memory-less. A system $S$, described by a density matrix $\rho_S \in \mathcal{B}(\mathscr{H}_S)$ (here $\mathscr{H}_S$ is the Hilbert space of the system), which follows a Markovian dynamic (of Hamiltonian $H$) under the influence of some bath $B$, evolves according to the Lindblad equation \cite{lindblad1976generators, breuer2002theory, alicki2019introduction, manzano2020short}
\be \label{lindbladeq}
    \frac{d \rho_S}{dt} = \mathcal{L}(\rho_S) = - i[H, \rho_S] + \mathcal{D}(\rho_S), \qquad \mathcal{D}(\rho_S) = \sum_i \g_i (L_i \rho_S L_i^\dag - \f12 \{ L_i^\dag L_i, \rho_S \}), \qquad \g_i \geq 0 \, ,
\ee
ensuring a (complete) positive and trace preserving dynamics. In \eqref{lindbladeq}, $\mathcal{L}$ is the \emph{Lindblad operator}, $\mathcal{D}(\rho)$ is called the \emph{dissipator} and encodes the non-unitary part of the underlying dynamics. It is expressed via some \emph{jump operators} $(L_i, L_i^\dag)$ and \emph{dissipation rates $\g_i$}, whose exact expression is phenomenological.\footnote{They depend on the system at hand. If we imagine the system $S$ to be a qubit weakly interacting with a bath at thermal equilibrium, the jump operators $L_i$ can be the Pauli ladder operators $L_{\pm} = \s_{\pm}$ associated to the qubit (plus eventually a dephasing term induced by the operator $L_z = \s_z$).}  The Lindblad equation is the generalization of the von Neumann equation\footnote{In the Schr\"odinger picture is reads
\begin{equation}
    \label{eq: von Neumann eq}
    i\frac{d}{d t}\rho_S = [H, \rho_S] \, .
\end{equation}} 
in the case where a system is driven by an environment to which it is weakly coupled, so that we can consider that its dynamics is effectively Markovian. It can be proven \cite{lindblad1976generators} that any dynamics that is complete positive and trace preserving \footnote{So that we map density matrices into density matrices. The notion of complete positivity is needed so that an induced dynamics on a subsystem preserves the positivity of the larger system.} (CPTP) and Markovian can be written in the Lindblad form \eqref{lindbladeq}.

However, in realistic physical situations, strict Markovianity is an idealization. Indeed, once the system $S$ and the bath $B$ begin to interact, correlations inevitably develop between them, regardless of their initial states. The assumption underlying the macroscopic derivation of the Lindblad equation is that such correlations are effectively “erased” by the bath after a characteristic time scale $\tau_B$ which can be interpreted as the bath correlation time. This time scale is typically of order $\frac{\hbar}{T_B}$, $T_B$ denoting the temperature of the bath \cite{breuer2002theory}.\footnote{This estimate applies provided the bath does not possess additional intrinsic scales, such as infrared or ultraviolet cutoffs.}
On time scales longer than $\tau_B$, the reservoir effectively loses memory of its past interactions by dispersing information into its infinitely many degrees of freedom, so that the reduced dynamics of the system becomes effectively Markovian. Consequently, the time variable $t$ appearing in the Lindblad equation~\eqref{lindbladeq} should not be identified with the microscopic time entering the von Neumann equation. Rather, it should be interpreted as a coarse-grained, mesoscopic time variable $\tau$.  As a result, the Lindblad equation can only be used to describe the dynamics of the system $S$ on time scales $t \gg \tau_B$, and any attempt to interpret it at shorter time scales is physically meaningless.\footnote{This is analogous to expecting the Navier–Stokes equations to remain valid below the atomic scale, where the notion of a continuum fluid description breaks down.} The joint evolution of the system $S$ and the environment $B$ is schematically represented by the evolution
\be
    \rho_{SB}(0) = \rho_S(0) \otimes \rho_B(0) \longrightarrow \rho_{SB}(\tau_B) = \rho_{S}(\tau_B) \otimes \rho_{B}(\tau_B) + \delta \chi \underset{t \gg \tau_B}{\longrightarrow} \rho_{SB}(t) \approx \rho_S(t) \otimes \rho_B(t)
\ee
where $\delta \chi$ represents the correlations induced between the system and the bath. \footnote{At this stage, a nonvanishing mutual information between the system and the bath is generated by the interaction. As a consequence, the second law may be locally violated due to information backflow into the system. This mechanism underlies Maxwell’s demon–type scenarios, in which available information is consumed to apparently violate the second law.}
This information is subsequently erased as it is dissipated into the infinitely many degrees of freedom of the bath,\footnote{\label{foot: interaction hamiltonian}More precisely, by $\rho_B$ we mean the state of the subset of bath degrees of freedom that effectively interact with and drive the system $S$. If instead one were to identify the bath with “the rest of the universe,” correlations would never truly disappear and no information would be lost. What is required is that the correlation functions of the bath observables $B_\alpha$ associated with these degrees of freedom decay on time scales $t \gg \tau_B$, namely $\langle B_\alpha(t) B_\beta(t + \Delta t) \rangle \underset{\Delta t \gg \tau_B}{\longrightarrow} 0$. In practice, these operators $B_\alpha$ appear in the interaction Hamiltonian between the system and the bath $H_{I} = \sum_\alpha S_\alpha \otimes B_\alpha$ where the $S_\alpha$ are operators acting on the system $S$.} and the reduced dynamics of the system becomes effectively Markovian on time scales larger than $\tau_B$.  

In our setting, the characteristic time scale $\tau_B = \kappa^{-1} \sim \frac{\hbar}{T_H}$ corresponds to the inverse Hawking temperature and plays the role of the reservoir correlation time. It is also of the order of the typical wavelength of Hawking radiation, and thus naturally characterizes the decay of correlations. Open quantum system theory then implies that the second law is expected to hold only on time scales much larger than the correlation time $\tau_B = \frac{\hbar}{T_B}$. Accordingly, we do not expect the dual generalized second law to be valid on time scales of order $\tau_D \sim \tau_B$, but only on longer, coarse-grained time scales $\tau_D \gg \tau_B$. This requirement is precisely captured by assumption  \eqref{assumpthor2draft2}.

At this point, it is useful to explain more precisely why this parallel can be drawn. The key observation is that, on null hypersurfaces, Markovianity is not merely an approximate property—it is exact, enforced by causality: once information crosses a causal horizon, it cannot return. Consequently, one would expect the effective correlation time to vanish, i.e. $\tau_B = 0$. Indeed, this is exactly what occurs on null hypersurfaces at finite distance, such as black hole horizons or the Rindler horizon. This is perfectly consistent with our formalism: an observer attempting to measure an operator on a future causal horizon would require infinite acceleration, corresponding to an infinite Unruh temperature. Likewise, the typical wavelength of particles observed exactly on a black hole horizon vanishes due to infinite blue-shift, so the correlation time is zero. Similar reasoning applies to any null hypersurface at finite distance. 

The situation is different at null infinity. Here, $\scri^+$ is a null hypersurface in the conformal spacetime, and the observers following its null geodesics are inertial. This explains why the natural time scale of the problem, the surface gravity $\kappa$, appears explicitly in formulas such as \eqref{kuzerodefdraft2} and \eqref{diffbondimassdraft2}, since we have chosen $u$ to be the affine inertial time. In contrast, it does not appear in the analogous expressions for finite horizons, such as \eqref{differenceareasde1draft2}. It is well known that the surface gravity is fundamentally associated with physics at infinity rather than on the horizon itself: it is defined by normalizing the asymptotic Killing vector so that its timelike component has unit norm at infinity. Consequently, it corresponds to the asymptotic generator of time translations.\footnote{Its Noether charge then corresponds to the usual notion of energy.} On the horizon, by contrast, the Killing field is null and cannot be normalized.

\subsection{Dual Generalized Second Law from Lindblad's equation}

In this subsection, we take a step back and analyze the formulas obtained in subsection \ref{subsec: reminder dual GSL} from a thermodynamic point of view. First, we try to understand why we find additional terms to the energy and entropy variations when we look at the monotonicity of the relative entropy to infer a second law.

The usual framework in the theory of open quantum systems is to work with a finite dimensional system interacting with an infinite bath. Then, as we saw in subsection \ref{subsec: time scales}, as long as the coupling $\g$ is sufficiently weak and that the time scales we are interested in are longer than the correlation time $\tau_B$, \footnote{In fact, the relevant condition to make it work is to have $\tau_B \gamma \ll 1$ since the "transition time" must be of the order $\g^{-1}$ .} the system's dynamics can be approximated by a Markovian evolution which translates to the Lindblad equation \eqref{lindbladeq} at the quantum level. However, in order to deduce the precise form of the jump operators, we need to get a minimum of knowledge on the state of the bath, since the latter induces the system's dynamics. Since the bath is infinite, we cannot associate a proper density matrix to it. However, the state of the bath is implicitly characterized by the correlation functions of the bath operators. It is quite standard to assume that the operators of the bath satisfy the Kubo-Martin-Schwinger (KMS) conditions \cite{kubo1957statistical, martin1959theory} 
\be \label{KMScond}
    \forall (A, B) \in \mathcal{A}_B, \qquad \forall t \in \mathbb{R}, \qquad \langle \alpha_{t} (A) B \rangle = \langle B \alpha_{t + i \beta} (A) \rangle
\ee
where $\mathcal{A}_B$ is the bath's algebra, $\alpha_t$ the modular flow (see below \eqref{eq: KMS condition draft 2}) and $\beta$ is the inverse bath temperature. However, even if \eqref{KMScond} corresponds to the usual "thermality" conditions, they do not mean that the bath is thermal in the usual thermodynamic sense. Typically, if the bath cannot only exchange energy but also particles with the system, the flow $\a_t$ will not be generated by the system's Hamiltonian, and will generically depend on the Lagrange multipliers associated to the conserved quantities ($\beta$ for the energy and the chemical potential $\mu$ for the number of particles). Indeed, if $\{ a_\om, a_\om^\dag \}_{\om \in \mathbb{R}}$ is the set of ladder operators associated to the bath, in such cases the KMS conditions will imply that
\be \label{KMSchemicalpotentails}
    \langle N_\om \rangle = \langle a_\om^\dag a_{\om} \rangle = \langle a_\om a_{\om}^\dag e^{-\beta (\om - \m)} \rangle = (\langle a_\om^\dag a_\om  \rangle + 1) e^{-\beta (\om - \m)} = (\langle N_\om  \rangle + 1) e^{-\beta (\om - \m)}
\ee
so that if $n_\om = \langle N_\om \rangle$ is the average number of particles, we have \footnote{In fact, one can obtain much more that only the average number of particles but the whole Boltzmann distribution by using the KMS conditions from the expectation value $\forall k \in \mathbb{N}, \langle N_\om^k \rangle = \langle a_\om N_\om^{k-1} a_\om^\dag e^{- \beta(\om - \m)} \rangle = \langle (N_\om + 1)^{k} e^{- \beta(\om - \m)} \rangle$ (the latter equality requires the identity $a_\om N_\om^k = (N_\om + 1)^k a_\om$ which can be easily proven by induction ), to deduce the set of equations 
\be
    \forall k \in \mathbb{N}, \quad \sum_{N_\om = 0}^{+ \infty} (N_\om + 1)^k (p( N_\om + 1) - e^{-\beta (\om - \m)} p( N_\om)) = 0
\ee
leading to the Boltzmann distribution.
}
\be
    \frac{n_\om + 1}{n_\om } = e^{\beta (\om - \m)}
\ee
which is a Bolse-Einstein distribution. In order to gain some intuition, it is insightful to write the formal density matrix associated to the bath as 
\be
    \rho_{B} = \frac{e^{- \beta (H_B - \sum_i \m_i N_i)}}{Z}, \qquad Z = \Tr{e^{- \beta (H_B - \sum_i \m_i N_i)}}
\ee
where $H_B$ is the bath's Hamiltonian and $\m_i$ are the chemical potentials associated to the species $i$ (whose number operator is of course $N_i$). Then the modular flow $\a_t$ can be written as 
\be
    \forall A \in \mathcal{A}_B, \quad \forall t \in \mathbb{R}, \qquad \a_t(A) = \rho^{it}_B \, A \, \rho^{-it}_B \, .
\ee
Similarly, in Tomita-Takesaki modular theory reviewed in subsection \ref{subsec: reminder algebra} (see also Appendix D of \cite{RBVilatte251}), any cyclic and separating state $\ket{\Psi}$ satisfies the KMS condition with respect to its own modular Hamiltonian  $K_{\Psi} = - \ln \Delta_\Psi$ (with inverse temperature $\beta = 1$), so that
\be \label{KMScond2}
    \forall (A, B) \in \mathcal{A}_B, \qquad \forall t \in \mathbb{R}, \qquad \bra{\Psi} \Delta^{i(t + i)}_\Psi A \Delta^{-i(t + i)}_\Psi  B \ket{\Psi} = \bra{\Psi} B \Delta^{it}_\Psi A \Delta^{-it}_\Psi \ket{\Psi}
\ee
and in particular, the (algebraic) vacuum state $\om$ from which the GNS Hilbert space $\mathscr{H}_\Om$ is built satisfies the KMS conditions with respect to its own modular Hamiltonian $K_{\Omega} = -\ln \Delta_\Om$. However, a generic quasifree state $\om$ is not thermal, since its modular Hamiltonian does not coincide with the Noether charge associated with time translations. We have seen that, when the Hartle–Hawking state $\om_H$ is restricted to the algebra $\mathcal{A}_0$, the modular Hamiltonian reduces precisely to the standard generator of time translations on $\scri^+_R$ (see \eqref{modhamnormordidraft2}), as shown in \cite{RBVilatte251}. By contrast, for the Minkowski vacuum $\om_M$, the modular Hamiltonian is given by the boost generator. For the $\kappa_l$-vacuum $\om_H^{\{ \kappa_l \}}$, the modular Hamiltonian differs from both cases and instead resembles that of a thermal bath at equilibrium with nonvanishing chemical potentials. From a physical perspective, selecting a vacuum state $\om$ on $\scri_R$ is therefore equivalent to choosing a specific reservoir--or infinite bath--in the framework of open quantum thermodynamics.  

We come now to the proof of the second law for open quantum systems undergoing a Markovian dynamics. The key ingredient is the fixed point of the dynamics induced by the Lindblad equation \eqref{lindbladeq}. However, it is easy to prove that if the two point functions of the observables of the environment \footnote{At least the ones that appear in the interaction Hamiltonian $H_I$ between the system and the infinite bath, see footnote \ref{foot: interaction hamiltonian}.} satisfy the generalized KMS conditions \eqref{KMScond}, the fixed point $\s \in \mathcal{B}(\mathscr{H}_S)$ will correspond to an equilibrium state relative to the environment. For instance, if the correlation functions of the reservoirs behave so that the KMS conditions \eqref{KMSchemicalpotentails} are satisfied, it is easy to show that the fixed point of the dynamics will be \footnote{Contrary to the environment, the system has a finite dimensional Hilbert space, so that its density matrix is well defined.} 
\be \label{fixedpointnonvmu}
    \s = \frac{e^{- \beta (H_S - \sum_i \m_i N_i)}}{Z}, \qquad Z = \Tr{e^{- \beta (H_S - \sum_i \m_i N_i)}}
\ee
where $H_S$ is the system's Hamiltonian and $N_i$ is the number operator, of quanta of the type $i$ that are exchanged between the system and the bath. Then, it is possible to prove that the eigenvalues of the Lindbladian appearing in \eqref{lindbladeq} are all strictly negative except one that is equal to zero, so that there is a unique fixed point of the dynamics.\footnote{Any CPTP map that is an automorphism admits at least one fixed point, but in general this fixed point is not unique. In order to obtain uniqueness, the dynamics should be able to "mix" all the different sectors of the Hilbert space, otherwise different sector could have their own fixed point.} The fact that the other eigenvalues are negative allows us to conclude that the state of the system will converge towards its fixed point $\rho \underset{t \rightarrow + \infty}{\longrightarrow} \s$. Calling $\lambda_1$ the smallest non-vanishing eigenvalue of the Lindbladian $\mathcal{L}$ (in absolute value), we can conclude from \eqref{lindbladeq} that for any $\rho \in \mathcal{B}(\mathscr{H}_S)$
\be \label{lamba1rel}
    \lvert \lvert \rho(t) - \s \lvert \lvert \leq e^{- \lambda_1 t} \lvert \lvert \rho(0) - \s \lvert \lvert
\ee
so that the system will "thermalize" (or relax) to $\sigma$ in a time $\tau_R \sim \frac{1}{\lambda_1}$. In order to study the thermodynamics of the process, we consider the relative entropy $S(\rho \lvert \lvert \s)$. Then, we use the monotonicity of the relative entropy for the CPTP map $e^{t \mathcal{L}}$ that is the propagator of the Lindbladian, and get 
\be
    \forall (s, t) \in \mathbb{R}^2_+, \qquad S(\rho(s) \lvert \lvert \s) \geq S(e^{t \mathcal{L}}\rho(s) \lvert \lvert \s) = S(\rho(s + t) \lvert \lvert \s) 
\ee
so that \footnote{Notice that the monotonicity of relative entropy is in general not enough to ensure thermalization. Since the relative entropy decreases and is positive, it must converge to some $L \geq 0$ at late times. However, thermalization to the fixed point is equivalent to prove that $L = 0$.} we can write 
\be \label{monotofrelentropylindblad}
    \Delta S(\rho) - \Delta \langle K \rangle_\rho \geq 0
\ee
where $K = - \ln{\s}$ is the modular Hamiltonian of the fixed point $\s$, and $S(\rho)$ is the von Neumann entropy of the state $\rho$. For instance, in the special case where the KMS conditions translate to \eqref{KMSchemicalpotentails}, so that \eqref{fixedpointnonvmu} is the fixed point of the dynamics, \eqref{monotofrelentropylindblad} becomes
\be \label{monotofrelentropylindblad2}
    \Delta S(\rho) - \beta (\Delta \langle H \rangle_\rho - \sum_i \m_i \Delta \langle N_i \rangle) \geq 0
\ee
or 
\be
    \Delta \mathcal{G} \leq 0, \qquad \mathcal{G} = \langle H \rangle_\rho - \sum_i \m_i \langle N_i \rangle_\rho - T_B S(\rho)
\ee
where $\mathcal{G}$ is the grand potential. Of course, if $\forall i, \m_i = 0$, then the grand potential reduces the free energy $\mathcal{F} = \langle H \rangle_\rho - T_B S(\rho)$. Therefore, the difference of relative entropy is indeed nothing more than the usual entropy production term that we have in ordinary thermodynamics. It is of course very similar to the equations of the previous sections \eqref{eq: dual GSL HH} or \eqref{grandpotential2var}. In addition, since the variation of internal energy is given by $\Delta E = \Delta \langle H \rangle_\rho$, and since the Clausius relation can be written as 
\be
    \Delta S - \frac{Q}{T_B} = S_c \geq 0
\ee
(with $S_c$ the created entropy), one can identify the heat and work fluxes 
\be
\label{eq: decompo work and Q}
    Q = \Delta \langle H \rangle_\rho - \sum_i \m_i \Delta \langle N_i \rangle_\rho, \qquad W = \sum_i \m_i \Delta \langle N_i \rangle_\rho
\ee
where $W$ is here the chemical work. Of course, if we have more constraints so that more charges could be exchanged between the system and the reservoir, we could have additional work sources on the system, like mechanical work $- p \text{d}V$ or electrical work $- \Phi \text{d} Q$.\footnote{Electrical work appears in gravity when considering the Reissner-Nordstr\"om black hole.} This relation between the terms related to the chemical potential and the notion of thermodynamic work is essential to show how work can be extracted from the $\kappa_l$-vacuum, see Section \ref{sec: work extraction}. Hence, the thermodynamics induced from the Lindblad equation in a system that weakly interacts with some reservoir in a generalized KMS state is very similar to the story told in Section \ref{sec: GSL proof}. We summarize the different points in the following Table \ref{tablethermo2}. 

\begin{table}\begin{center}\begin{spacing}{1.3} 
\begin{tabular}{|l|l|l|}
\hline \emph{} & \emph{Quantum thermodynamics} & \emph{At future null infinity $\scri^+$} \\\hline
System & Our physical system & Excitations of the quantum field  \\
Reservoir & Infinite bath at equilibrium
&  Boundary conditions at late time  \\
Time scale & Coarse-grained time $t \gg \tau_B \sim \frac{1}{T}$ & Coarse-grained time $t \gg \tau_B \sim \frac{1}{T_H}$ \\
State invariant under the dynamics & The fixed point & The vacuum state $\om$ \\
Reference state in the relative entropy & The fixed point & The vacuum state $\om$ \\
Entropy production & 
Difference of relative entropies & Difference of relative entropies \\
Relaxation to equilibrium & Towards the fixed point & Towards the vacuum state 
\\\hline
\end{tabular}\end{spacing}\end{center}\vspace{-1cm}
\caption{\label{tablethermo2}\emph{\small{ The physics of a small system weakly coupled to a large reservoir in equilibrium undergoes a Markovian dynamic (driven by the Lindblad equation), exactly as the evolution of quantum fields defined on the non-expanding null hypersurface $\scri_R$, where by ``evolution" we mean restriction of the state to smaller and smaller subalgebras $\mathcal{A}_i^\scri$ of observables on $\scri_R$. }}}
\end{table}

The key point is that choosing a quasifree state $\om$—which, via the GNS construction, fixes a corresponding Hilbert space—is equivalent to selecting a reservoir that drives the system toward a specific equilibrium state. However, the reasons why the system equilibrates in the two cases are quite different. In a Hilbert space $\mathscr{H}_\om$, it is because any state $\ket{\Psi}$ will look like the vacuum state $\om$ if one sufficiently restricts the algebra of observables. Indeed, since a state $\ket{\Psi} \in \mathscr{H}_\om$ is (a linear combination of) local excitations of the field on top of the vacuum state, these local excitations will not be visible from the subalgebra $\mathcal{A}_i$, if the cut $u_i$ is taken at sufficiently late time. For instance, a coherent state 
\be
    \ket{\Psi} = e^{i \pi(f)} \ket{\Om} \in  \mathscr{H}_\om
\ee
generated by a test function $f$ with compact support on $I \subset \scri_R$ will look like the vacuum with respect to the algebra $\mathcal{A}_i$ if $I \cap (u_i, + \infty) = \emptyset$. Consequently, since $ e^{i \pi(f)}$ is a unitary operator, the relative entropy between $\ket{\Psi}$ and $\ket{\Om}$ will vanish when restricted to the algebra $\mathcal{A}_i$, as explained below \eqref{eq: relative entropy draft 2}. However, there does not exist a time scale comparable to $\tau_R \sim \lambda_1^{-1}$ as in \eqref{lamba1rel}. We can understand it because here the Hilbert space is infinite dimensional, so there is no equivalent to a strictly non vanishing lower bound for the eigenvalues of $\mathcal{L}$. Therefore, in general we must take $\tau_R = + \infty$ i.e. the process we just described happens strictly speaking only at ``infinite" late times. We should not forget that we mean very different things when we talk about the "evolution" of the quantum system in both cases (algebraic and thermodynamic). Indeed, we \textit{define} the state of the quantum field on $\scri_R$, so that its specification is \textit{kinematical}, and by evolution we only mean to restrict some initial data to a a smaller subset of data, while actually the Lindblad equation talks about a \textit{dynamical} evolution of the quantum state. However, the two concepts coincide on a null hypersurface: they are the limits of both spacelike and timelike hypersurfaces, so that we can define quantum states and an Hilbert space on them, and still see the (kinematical) restriction of these states to a smaller subalgebra at late time as a dynamical evolution. This is why there is such an interesting interplay between thermodynamics and the physics of quantum field on null hypersurfaces. The choice of the vacuum state fixes the infrared behavior of the field on the null hypersurface. However, contrary to what happens on a Cauchy surface, in the case at hand, infrared behavior means \textit{late time} behavior, so that the field relaxed to the vacuum state at late time on the null hypersurface. The different inequivalent vacuum states offer different thermodynamic stories to the quantum field. 

To conclude this subsection, we would like to come back to what we identified as the "system" and what we identified as the "reservoir" for the quantum fields at $\scri^+_R$. As we have emphasized, the choice of Hilbert space, or equivalently the choice of vacuum state $\om$, is analogous to a choice of reservoir. Therefore, what we call the system here is just made out of the excitations of the quantum fields on top of the vacuum state $\ket{\Om}$, the latter being fixed by the late time boundary conditions. In the vacuum state $\ket{\Om}$ the quantum field behaves exactly as a reservoir would behave, since this state is stationary. Consequently, there is no entropy production and the vacuum state is in thermodynamic equilibrium. However, as we also emphasized in the previous subsections, in general, even in the vacuum state the charges are not conserved, so that we expect a constant flux of particles, energy and entropy. However, since the vacuum state is stationary, the variation of thermodynamic quantities is made without any entropy production ($S_c = 0$) so that the Clausius relation
\be \label{Clausiusrelation}
    \Delta S = \frac{Q}{T}
\ee
is always satisfied. In usual thermodynamics, an infinite bath  obeys the Clausius relation \eqref{Clausiusrelation} since it is always at equilibrium and there are no entropy production term. Remember that in the vacuum state, the charges, like the total energy, are strictly speaking infinite, so there can be constant energy and entropic fluxes without changing the state. These are exactly the same assumptions that are made for a bath at equilibrium in thermodynamics, which always satisfies some kind of generalized KMS conditions even if it can exchange energy and entropy with the system. Indeed, if $A$ is the system (we change the latter wrt subsection \ref{subsec: time scales} to avoid confusion with the entropy) and $B$ is the bath, the usual thermodynamic relations tell us that 
\be
    \Delta S_A = \frac{Q_A}{T} + S_c, \qquad \Delta S_B = \frac{Q_B}{T} 
\ee
where $T$ is the temperature of the bath. In addition, conservation of energy between the bath and the system implies that $Q_A + Q_B = 0$, so that
\be
    \Delta S_A + \Delta S_B = S_c \geq 0
\ee
for the closed system. For quantum fields on $\scri^+_R$, $Q_B$ is the constant heat flux of the vacuum state (equal to $\Delta M_\Om$ for instance if all the chemical potentials vanish), while $Q_A$ is the heat flux associated to the excitations on top of the vacuum, i.e. are related to the one-sided modular Hamiltonian that vanishes in the vacuum $\om$. However, here we do not have $Q_A + Q_B = 0$, since the system is open and therefore energy, particles and entropy can flow through $\scri^+_R$, so the total charges (the Bondi mass, the number of particles, etc...) are not conserved, and it is the reason why one has to deal with other thermodynamic potentials than the entropy.


\section{Work extraction from a non-rotating black hole}
\label{sec: work extraction}

In this Section we explain how we could interpret the additional ``chemical potential" terms appearing in the formula \eqref{grandpotential2var} as work, so that from the Hilbert space $\mathscr{H}_{\Om_H^{\{ \kappa_l \}}}$, one can design an engine which runs based on the fact that the vacuum $\ket{\Om_H^{\{ \kappa_l \}}}$ is not thermal. In order to establish the latter, we will discuss the notion of pasive states and work out an explicit example of how we could extract work using the Brunner-Linder-Popescu-Skrzypczyc (BLPS) engine \cite{brunner2012virtual}.  

\subsubsection*{Passive states}

In order to understand why we can extract work from the $\kappa_l$-vacuum, it is essential to define first what is a passive state \cite{pusz1978passive, alicki2019introduction}. 

\vspace{10pt}

\textbf{Definition 1.} A passive state is a quantum state $\rho$ such that any unitary transformation on it can only increase its internal energy, i.e for any unitary $U \in \mathcal{B}(\mathscr{H}_\Omega)$
\be \label{passivestate1}
    \tr{(H U \rho U^\dag)} \geq \tr{(H \rho)} \, .
\ee
From this definition, it is straightforward to show that an alternative characterization of a passive state would be that the population of the state in the energy eigenbasis decreases if the energy increases, i.e.
\be \label{passivestate2}
    \forall (E, E'), \quad E' > E \Rightarrow p(E') < p(E),
\ee
since a unitary transformation mixes the eigenvectors while keeping the same eigenvalues. The passive states are very important in quantum thermodynamics, because it is impossible to extract work from them using a unitary operation. Otherwise, if the state is not passive, it is in principle possible to extract work. A standard example of a passive state is a thermal state, which obviously satisfies the two equivalent conditions \eqref{passivestate1} and \eqref{passivestate2}. Indeed, it is well-known that it is impossible to extract work from a \textit{single} thermal state, as it was even one of the earlier formulations of the second law of thermodynamics (due to Kelvin and Planck).

From this new perspective, it is interesting to understand which are the passive states among the different vacua we considered in Section \ref{sec: GSL proof}. It is quite clear that the Hartle-Hawking state $\ket{\Om_H}$ (restricted to any algebra $\mathcal{A}_i^\scri$) is a passive state, since it is thermal. Same for the $L$-vacuum which is exactly thermal up to the cutoff $L$, and finally coincides with Minkowski. However, the $\kappa_l$-vacuum is certainly not passive in general because of the presence of the chemical potentials $\mu_{\omega l}$ (see \eqref{eq: relation mu kappal}). Indeed, one can always find (many) pairs of triplets $(N_{\om l m}, \om, l)$ and $(N_{\om' l' m'}, \om', l')$ so that
\be
    e^{- \frac{2 \pi N_{\om l m} \om}{\kappa_l}} \leq e^{- \frac{2 \pi N_{\om' l' m'} \om'}{\kappa_{l'}}}, \qquad N_{\om l m} \om \leq N_{\om' l' m'} \om'
\ee
as long as $\kappa_{l'} \neq \kappa_l$, which contradicts \eqref{passivestate2}. It is because, precisely, the $\kappa_l$-vacuum is not thermal with respect to the generator of asymptotic translations \footnote{More precisely, with respect to $H = \int_{\mathcal{D}_i^\scri} (1 - e^{- \kappa(u - u_i)}) T_{uu} \text{d}u \wedge \epsilon_S$ that is the correct one-sided modular Hamiltonian for the algebra $\mathcal{A}_i^\scri$.} when restricted to the algebra $\mathcal{A}_i^\scri$. Likewise, as we will explain in Section \ref{sec: vac unruh}, the Unruh vacuum $\om_U$ is not a passive state when restricted to the algebra $\mathcal{A}_i^\scri$. Again, it is because the Unruh vacuum state $\om_U$, although thermal if restricted to the white hole horizon $\mathcal{H}_R^{-}$ is not thermal on $\scri^+_R$ since the potential barrier scatters the particles emitted from $\mathcal{H}^-_R$. 

\subsubsection*{The BLPS engine}

In order to illustrate the previous statements, it is useful to explicitly show how we can extract work from the $\kappa_l$-vacuum state, since we claim it should be feasible. An interesting illustration of how it could work is to study the BLPS engine \cite{brunner2012virtual}.

\begin{figure}[ht]
\begin{center}
\includegraphics[scale=0.2]{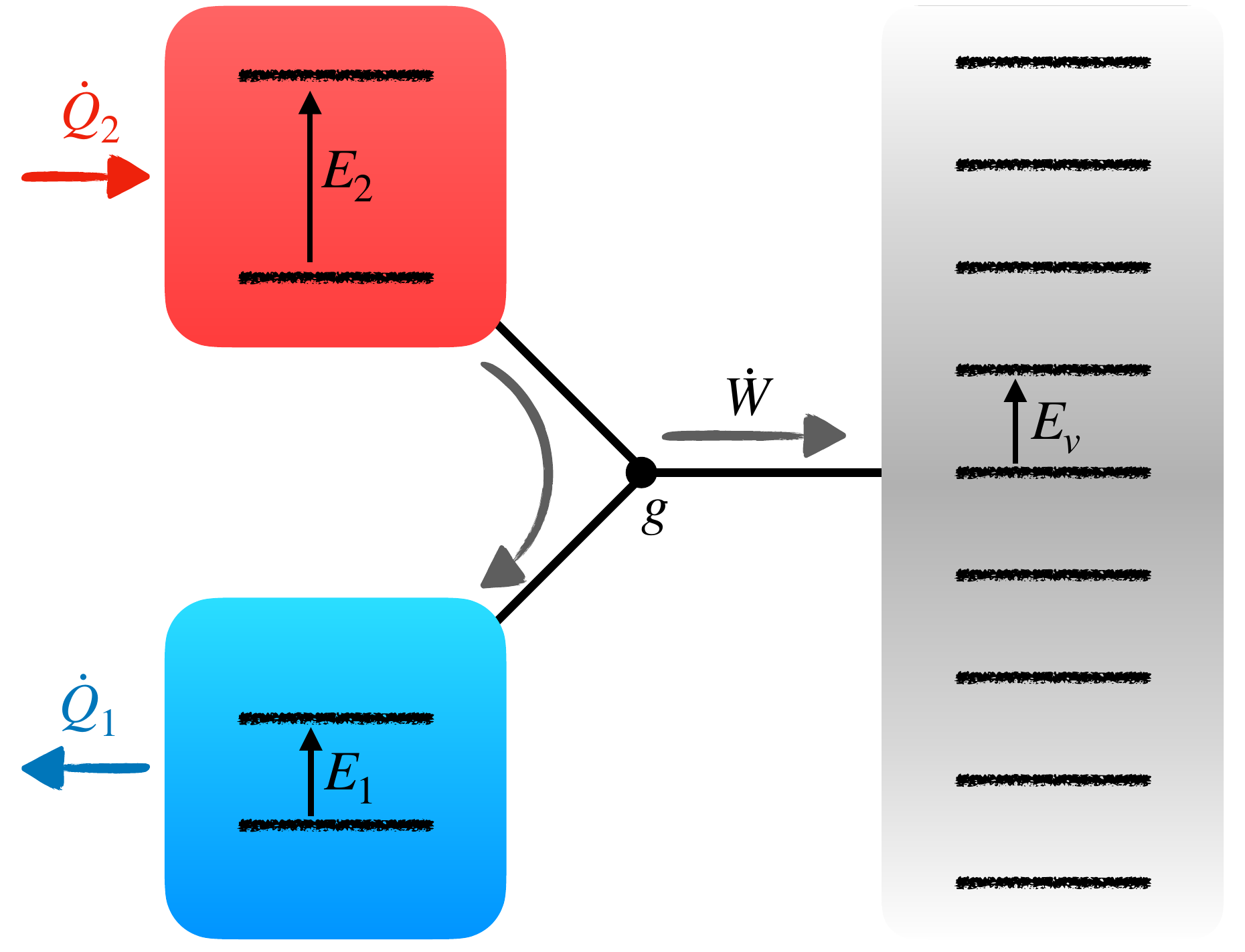}
\caption{A picture depicting the BLPS engine. Because of the population inversion, we are able to extract some energy (heat) $Q_2$ from the hot temperature bath so that we can perform some work $W \leq 0$ to lift a load on an energy ladder. Figure taken from \cite{Rignon-Bret:2020nmt}, written by one of us.}
\label{BLPSengine}
\end{center}
\end{figure}

\begin{figure}[ht]
\begin{center}
\includegraphics[scale=0.3]{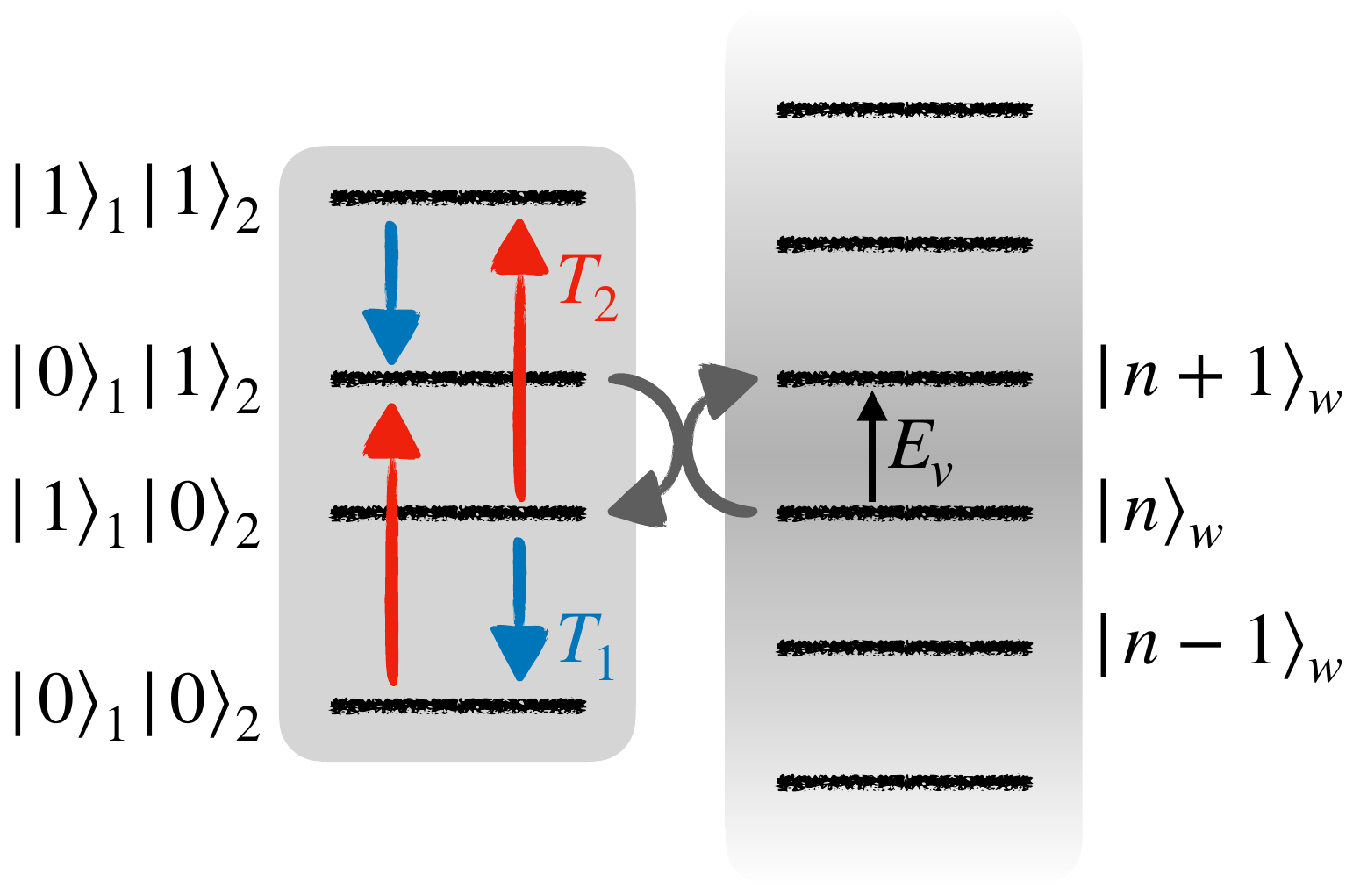}
\caption{Here we represented the eigenbasis of the tensor space spanned by our two qubits. We depicted the jumps induced by the reservoirs (in blue and red) on the composite qubit and the transition to a higher energy eigenstate of the ladder. Figure taken from \cite{Rignon-Bret:2020nmt}, written by one of us.}
\label{Virtualqubit}
\end{center}
\end{figure}

This engine works by taking two qubits made of two energy eigenstates (with different energies) $(\ket{0}_i, \ket{1}_i)$, $i = 1,2$, coupled to two thermal baths at different temperature $T_1$ and $T_2$, with $T_2 > T_1$. Let us assume that the energy gap between the ground states $\ket{0}_1$ (resp. $\ket{0}_2$) and the excited states $\ket{1}_1$ (resp. $\ket{1}_2$) of the qubit of the cold (resp. hot) bath is $E_1$ (resp. $E_2$), and that $E_2 \geq E_1$. The four energy eigenstates states $(\ket{0}_1, \ket{0}_2, \ket{1}_1, \ket{1}_2)$ of the two qubits span a four dimensional Hilbert space $(\ket{00},  \ket{10}, \ket{01}, \ket{11})$ by taking the tensor product of the Hilbert spaces of the two separate qubits. Since the cold (resp. hot) bath is in thermal equilibrium at the temperature $T_1$ (resp. $T_2$), we have that 
\be \label{probabilitiesstatesblps}
    p(\ket{10}) = e^{- \frac{E_1}{T_1}} p(\ket{00}), \qquad p(\ket{01}) = e^{- \frac{E_2}{T_2}} p(\ket{00}), \qquad p(\ket{11}) = e^{- \left(\frac{E_1}{T_1} + \frac{E_2}{T_2}\right)} p(\ket{00})
\ee
and the stationary state of the joint qubits \eqref{probabilitiesstatesblps} is not passive if and only if 
\be
    \frac{E_2}{T_2} - \frac{E_1}{T_1} \leq 0
\ee
so that we have a population inversion, i.e.
\be \label{populationinversion}
     p(\ket{10}) \leq p(\ket{01}), \qquad E_2 \geq E_1
\ee
Of course, it is impossible to satisfy \eqref{populationinversion} if $T_1 = T_2$ but feasible if one can describe the environment as a bunch of thermal baths at different temperatures. However, if \eqref{populationinversion} is satisfied, we can use this non-equilibrium configuration to lift a load on an energy ladder. Indeed, if we assume that the energy gap on the ladder is given by 
\be
    E_v = E_2 - E_1 \geq 0
\ee
then it is enough to couple the \textit{virtual qubit} spanned by the states $(\ket{10},\ket{01})$ to the eigenstates of the ladder through the (pertrubative) interaction Hamiltonian (see Figure \ref{BLPSengine})
\be \label{InteractionAHmiltonian}
    H_I = g \sum_{n} \ket{0 1} \ket{n} \bra{1 0} \bra{n + 1} + \text{c.c}, \qquad g \ll E_1, E_2
\ee
where the the states $\ket{n}$ are the energy eigenstates of the ladder with eigenvalue $E_n = n E_v = n(E_2 - E_1)$. Now, the key point is that the population inversion \eqref{populationinversion} of the virtual qubit runs the engine so that the load actually \textit{climbs} the energy ladder instead of lowering it. Indeed, the transition $\ket{01} \rightarrow \ket{10}$ is more likely than the inverse transition $\ket{10} \rightarrow \ket{01}$ since the state $\ket{01}$ is more populated than the state $\ket{10}$, so that the interaction term \eqref{InteractionAHmiltonian} induces the transitions $\ket{n} \rightarrow \ket{n + 1}$ more often than their counterpart $\ket{n+1} \rightarrow \ket{n}$, see Figure \ref{Virtualqubit}. In the end, on average, the load is able to climb the ladder so that we have extracted work form the population inversion \eqref{populationinversion}.

The exact dynamics of the engine is needed in order to compute the associated work (per unit of time) $\dot{W}$. However, one can give an upper bound for the efficiency of such an engine from basic thermodynamics 
\be
    \frac{\lvert \dot{W} \lvert}{\dot{Q}_2} \leq \eta_C = 1 - \frac{T_1}{T_2}
\ee
where $\eta_C$ is the famous Carnot's efficiency for an engine.

\subsubsection*{Application to $\scri^+_R$}

Likewise, the fact that the vacuum $\ket{\Om_H^{\{ \kappa_l \}}}$ is not a thermal state allows us to lift a load on an energy ladder exactly as the BLPS engine does. Indeed, when restricted to the algebra of observables $\mathcal{A}_i^\scri$, we saw that the $\kappa_l$-vacuum was a highly mixed state which satisfies the KMS conditions with different temperatures depending on the mode $l$. From the latter we get
\be
    \langle N_{\om l} \rangle_{\Om_H^{\{ \kappa_l \}}} = \langle a_{\om l}^\dag a_{\om l} \rangle_{\Om_H^{\{ \kappa_l \}}} = \langle a_{\om l} a_{\om l}^\dag e^{-\frac{2 \pi}{\kappa_l} \om} \rangle_{\Om_H^{\{ \kappa_l \}}} \Rightarrow \frac{ \langle N_{\om l} \rangle_{\Om_H^{\{ \kappa_l \}}} + 1}{ \langle N_{\om l} \rangle_{\Om_H^{\{ \kappa_l \}}}} = e^{\frac{2 \pi}{\kappa_l} \om}
\ee
so that the $\kappa_l$-vacuum is obviously not a passive state, since if $\kappa_l < \kappa$ (obtained when $l=0$) one can always find $\om' > \om$ so that 
\be \label{popinvqubitlvac}
    \frac{\om}{\kappa} - \frac{\om'}{\kappa_l} \leq 0 \, ,
\ee
as $\omega$ and $\omega'$ span a continuum set of values. Therefore, to operate the BLPS engine, one must identify two qubits with energy gaps $\om$ and $\om'$, and couple them to two appropriately chosen, distinct sectors $l$ of the outgoing radiation in the $\kappa_l$-vacuum. This arrangement ensures that the qubits thermalize at the respective temperatures $T_H = \frac{\kappa}{2 \pi}$ and $T_l = \frac{\kappa_l}{2 \pi}$.\

Then, we couple the two eigenstates $\ket{01}$ and $\ket{10}$ of the two qubits to an energy ladder using the perturbative interaction Hamiltonian \eqref{InteractionAHmiltonian}, exactly as we did in the previous paragraph. Consequently, we are indeed able to lift the load and so we can indeed extract work from the $\kappa_l$-vacuum state as long as the condition \eqref{popinvqubitlvac} is satisfied, as we advertised. It might be quite surprising at first because we have already highlighted the similarities between the $\kappa_l$ vacuum and the Unruh vacuum \cite{Unruh:1976db}, the latter being also not a passive state when restricted to $\scri^+_R$. One can therefore conclude that we can extract some work from a Schwarzschild black hole in the Unruh vacuum, while it seems counter-intuitive if one only has the Penrose process \cite{penrose1971extraction} in mind. However, we should remember that the Penrose process is a classical process, while here we are saying that we can extract work from the non thermality of the Hawking radiation at null infinity, which does not exist classically. Similarly, the first laws of black hole physics (the equilibrium version as well as the physical process version) are \textit{classical} laws, and therefore are referred as first laws of black hole mechanics. They cannot tell us about the decomposition into work and heat fluxes at null infinity of quantum processes at $\scri^+_R$. Instead, we should refer to operational definitions of work and heat, as we did in \eqref{eq: decompo work and Q}.

To find the maximal amount of work that can be extracted we start by using reversibility which tells us that
\be
    \frac{Q_1}{T_1} = \frac{Q_2}{T_2} \Rightarrow \frac{\om'}{T_l} = \frac{\om}{T_H}
\ee
so\footnote{It might seem paradoxical that the reversibility (and so the most efficient thermalization) happens when \eqref{popinvqubitlvac} is saturated, while we clearly need the population inversion to lift the load on the energy ladder. It is because it is just a limit for which the process becomes extremely slow. However, we should not confuse thermodynamics and kinetics. The thermodynamic transformation that consumes the least amount of resources is always the reversible one.} that 
\be
    \lvert W_{\text{max}} \lvert = \eta_C Q_2 = (1 - \frac{T_l}{T_H}) \om = (\frac{T_H}{T_l} - 1) \om' = - \m_{\om' l} 
\ee
is equal to the chemical potential associated to the mode $(\om', l)$. So, indeed, we see that we can interpret the term $- \m_{\om l} \langle n_{\om l m} \rangle$ in \eqref{grandpotential2var} as a resource that we can consume in order to extract work, for instance, as showed in this Section, to lift a weight. 


\section{Thermodynamics in the Unruh vacuum: Schwarzschild and Kerr}
\label{sec: vac unruh}

The vacuum states that we discussed in the previous sections were states defined on an asymptotic algebra of observables at null infinity. However, none of the states introduced so far would correspond to the restriction of the Unruh vacuum. Nevertheless, the Unruh vacuum is expected to be the quantum state of the fields at late time after a real black hole collapse. Hence we have a very good physical motivation to look at quantum states in the GNS representation of the Unruh vacuum and seek the formulation of the dual GSL in this case.

\subsection{Construction of the Unruh vacuum}
\label{subsec: construction Unruh}

The Unruh vacuum is constructed as a joint vacuum state on $\mathcal{H}_R$ and $\scri^-_L$. Since it can be shown that some components of the stress energy tensor blow up on the right horizon $\mathcal{H}_R$, the Unruh vacuum is usually only defined on the black hole region $III$ and the exterior region $I$. On the horizon $\mathcal{H}_R$, it is annihilated by the set of operators
\be
    a(f) = \left\{ 2 i \int_{\mathcal{H}_R} \bar{f} \pi \, \eps_{\mathcal{H}_R} \quad \lvert \quad f = \sum_{lm}   Y_m^l(x^A) \int_0^{+ \infty} a_{\Om l m}  e^{- i \Om \tilde{U}} \text{d}\Om \right\}
\ee
where $\pi = \p_{\tilde{U}} \phi$ is the momentum operator on the horizon, and $\tilde{U}$ the affine Kruskal coordinate on the horizon. In particular, in the Unruh vacuum $\om_U$ the two-point function of the momenta on the horizon $\mathcal{H}_R$ is given by \cite{kay1991theorems} 
\be \label{2ptfuncUn}
    \om_U \left(\p_{\tilde{U}} \phi(\tilde{U}_1, x^A_1) \p_{\tilde{U}} \phi(\tilde{U}_2, x^A_2)\right) = - \frac{1}{4 \pi} \frac{\delta^2(x^A_1 - x^A_2)}{(\tilde{U}_1 - \tilde{U}_2)^2}
\ee
which is invariant with respect to Mobius transformations of the coordinate $\tilde U$ (see also \cite{Wall:2011hj}). Hence, the restriction of the Unruh vacuum to the algebra of observables $\mathcal{A}^{\mathcal{H}_R}$ on the right horizon $\mathcal{H}_R$ is equal to the restriction of the Hartle-Hawking vacuum on $\mathcal{A}^{\mathcal{H}_R}$. Therefore, $\om_U$ satisfies the KMS conditions with respect to the algebra of observables on $\mathcal{H}_R^-$ (or equivalently $\mathcal{H}_R^+$) and the modular Hamiltonian is the local boost operator\footnote{As $\tilde U$ is affine and inertial on the right horizon the covariant and normal-ordered stress tensor coincides in the Unruh vacuum.}
\be
    K_{\Om_U}^{\mathcal{A}^{\mathcal{H}_R^-}} = - \ln{\Delta_{\Om_U}^{\mathcal{A}^{\mathcal{H}_R^-}}} =  2 \pi \int_{\mathcal{H}_R^-} \tilde{U} T_{\tilde{U} \tilde{U}} \text{d} \tilde{U} \wedge \eps_S \, .
\ee 
We are interested in the subregion of $\mathcal{H}^-_R$ (at $\tilde{V} = 0$), $\mathcal{D}_i^{\mathcal{H}_R} := (\tilde{U}_i, 0) \times S^2$ with $\tilde U_i < 0$. Note that this region of the horizon $\mathcal{H}^-_R$ directly faces the region $\mathcal{D}_i^\scri = (U_i, + \infty) \times S^2$ on $\scri^+_R$ (located at $\tilde{V} = + \infty$) with $U = -\frac{1}{\tilde U}$. Then, from Mobius invariance of the Unruh vacuum on $\mathcal{H}_R$, we deduce that $\om_U$ also satisfies the KMS conditions with respect to the algebra $\mathcal{A}_i^{\mathcal{H}_R^-}$ associated to the region $\mathcal{D}_i^{\mathcal{H}_R^-}$ on $\mathcal{H}_R^-$ with a modular Hamiltonian
\be \label{modhamiltonianuihorr}
    K_{\Om_U}^{\mathcal{A}_i^{\mathcal{H}_R^-}} = - \ln{\Delta_{\Om_U}^{\mathcal{A}_i^{\mathcal{H}_R^-}}} = 2 \pi \int_{\mathcal{D}_i^{\mathcal{H}_R^-}} (U - U_i) T_{UU} \text{d}U \wedge \eps_S \, .
\ee
use $\tilde{U} \rightarrow -\frac{1}{\tilde{U}} + \frac{1}{\tilde{U}_i} = -\frac{1}{\tilde{U}} - U_i$. Then, since $\om_U$ is thermal with respect to the modular Hamiltonian \eqref{modhamiltonianuihorr}, some outgoing particles are emitted with a thermal spectrum from the region $\mathcal{D}_i^{\mathcal{H}_R^-}$ of the past horizon $\mathcal{H}^-_R$ towards the exterior region $I$. Defining the Unruh vacuum directly on the algebra of observables at null infinity is a hard task that we bypass in this work by considering instead the propagation of the thermal spectrum emitted at $\mathcal{H}_R^-$ in the exterior region $I$, and by analyzing which part of it is actually received at $\scri_R^+$.

If we take $U_i = 0$, then $\tilde U_i = -\infty$ so $\mathcal{D}_i^{\mathcal{H}_R^-} := \mathcal{D}_{\infty}^{\mathcal{H}_R^-} = \mathcal{H}_R^-$, and the outgoing modes are thermal with respect to the Killing time $\p_u = \kappa U \p_U$.\footnote{Recall the expression of the Killing field of the exterior region $I$ displayed in \eqref{timelikekilling draft2}. On $\mathcal{H}_R^-$ it reads $\xi = -\kappa \tilde U \p_{\tilde U} = \kappa U \p_U$.} Therefore, the spectrum at $\scri^+_R$ can be computed using the transmission coefficients $t_{\om l}$ for any mode $\frac{Y^l_m}{\sqrt{4 \pi \om}} e^{- i \om u}$ defined on the horizon $\mathcal{H}_R^-$, where $\om$ is the Killing frequency conjugated to the Killing time. Indeed, since the Killing energy (and so the Killing frequency) is conserved while the mode propagates in region $I$, the problem is analogous to find the \textit{out} quantum state if we have a potential barrier with transmission coefficient $t_{\om l}$, while the \textit{in} state is a thermal state at temperature $T_H = \frac{\kappa}{2 \pi}$. Therefore, we deduce that the restriction of the Unruh vacuum at $\scri^+_R$ satisfies the following KMS conditions (valid for all $m$ given a fixed $l$)
\be \label{KMScondunruhinf}
    n_{\om l m} = \bra{\Om_U} a_{\om l m}^\dag a_{\om l m} \ket{\Om_U} = \bra{\Om_U} a_{\om l m} a_{\om l m}^\dag \ket{\Om_U} e^{-\beta_H(\om - \m_{\om l})} \Rightarrow \frac{n_{\om l m} + 1}{n_{\om l m}} = e^{\beta_H(\om - \m_{\om l})}
\ee
where
\be \label{chemicalpot}
    \m_{\om l} = T_H \ln{\frac{\lvert t_{\om l} \lvert^2}{1 - (1 - \lvert t_{\om l} \lvert^2)e^{- \beta_H \om}}}
\ee
and $\m_{\om l}$ is the chemical potential introduced in \cite{ARB24}. Indeed, formally, the one-sided modular operator of the Unruh vacuum associated to the algebra $\mathcal{A}_0^\scri := \mathcal{A}^{\scri_R^+}$ is given by 
\be \label{formalexpression}
    \s^{{\mathcal{A}_0^{\scri}}} = \sum_{\{ N_{\om l m} \}}\prod_{\om > 0, lm} e^{- \beta_H N_{\om l m} (\om - \m_{\om l})} \ket{N_{\om l m}} \bra{N_{\om l m}}
\ee
where the state $\s$ is not normalized and can be obtained directly by computing the ratio of particles from each mode thermally populated at the Hawking temperature on the past horizon $\mathcal{H}_R^-$ that have been able to cross the potential barrier (see \cite{ARB24} for a detailed computation). 

However, in order to apply our previous arguments of Section \ref{sec: GSL proof} (and detailed in \cite{RBVilatte251}) in the case of the Unruh vacuum, we want to compute the restriction of the latter to the subalgebra $\mathcal{A}_i^\scri$ of $\mathcal{A}_0^\scri$, i.e. the algebra of observables in the region $\mathcal{D}_i^\scri = (U_i, + \infty) \times S^2$ on $\scri^+_R$. In order to do this, we must be able to find the restriction of $\omega_U$ on the future event horizon $\mathcal{H}_L^+$ (with algebra $\mathcal{A}^{\mathcal{H}_L^+}$) and then the restriction of $\om_U$ on the algebra of observables $\mathcal{A}_i^{\mathcal{H}_R^-}$. The latter has already been worked out, and we saw that it was a thermal state with respect to the modular Hamiltonian \eqref{modhamiltonianuihorr}, with conjugated time 
\be \label{barutimeui}
    \bar{u} = \kappa^{-1}\ln{(U - U_i)} = u + \kappa^{-1} \ln{(1 - e^{- \kappa(u - u_i)})}
\ee
while the modular Hamiltonian associated to $\om_U$ on the algebra $\mathcal{A}^{\mathcal{H}_L^+}$ can be computed similarly to the asymptotic one \eqref{formalexpression} since we just have to compute the ratio of particles that have been reflected back towards the future horizon $\mathcal{H}^+_L$ knowing the initial thermal distribution on $\mathcal{H}_R^-$. For that we use the reflection coefficient $r_{\omega l}$, related to the transmission coefficient $t_{\om l}$ through the conservation equation
\be \label{rtconseq}
    \lvert r_{\om l} \lvert^2 + \lvert t_{\om l} \lvert^2 = 1 \, .
\ee
Knowing the restriction of the Unruh vacuum of a massless field on $\mathcal{D}_{i}^{\mathcal{H}_R^-}$ and $\mathcal{H}_L^+$ is sufficient to deduce its restriction on $\mathcal{D}_i^\scri$. However, even if in the Unruh vacuum one has an outgoing flux of thermal particles which are emitted from the region $\mathcal{D}_{i}^{\mathcal{H}_R^-}$, we cannot use \emph{a priori} the coefficients $(r_{\om l }, t_{\om l})$ to obtain the spectrum on $\mathcal{D}_i^{\mathcal{H}_R^-}$, because the frequencies appearing in $(r_{\om l }, t_{\om l})$ are the Killing frequencies conjugated to the Killing time $u$, while the Unruh vacuum is thermal on $\mathcal{D}_i^\scri$ with respect to the modular Hamiltonian associated to the time \eqref{barutimeui}, that is rigorously equal to the Killing time only if one takes $u_i = - \infty$ (i.e when $\mathcal{D}_{i}^{\mathcal{H}_R^-} = \mathcal{D}_{\infty}^{\mathcal{H}_R^-} = \mathcal{H}_R^-$). Nevertheless, we can decompose the modes $b_{\bar{\om} l m} = \frac{1}{\sqrt{4 \pi \bar{\om}}} Y_m^l e^{- i \bar{\om} \bar{u}}$ (vanishing for $u \leq u_i$) in the basis of modes $b_{\om l m} = \frac{1}{\sqrt{4 \pi \om}} Y_m^l e^{- i \om u}$ using the Klein-Gordon product
\begin{align}
    (b_{\om l m}, b_{\bar{\om} l' m'}) &= -2 i \int_{\mathcal{H}_R^-} b_{\bar{\om} l' m'} \p_u \bar{b}_{\om l m} \text{d}u \wedge \eps_S = \delta_{ll'} \delta_{mm'} \frac{1}{2 \pi} \int_{u_i}^{+ \infty} \sqrt{\frac{\omega}{\bar{\omega}}} e^{- i \bar{\om} \bar{u}}  e^{i \om u} \text{d}u \nn \\
    &= \frac{\delta_{ll'} \delta_{mm'}}{2 \pi} \int_{u_i}^{+ \infty} \sqrt{\frac{\omega}{\bar{\omega}}} e^{i(\om - \bar{\om})u}(1 - e^{-\kappa(u - u_i)})^{- i \kappa^{-1} \bar{\om}} \text{d}u \nn \\
    &= -\frac{\delta_{ll'} \delta_{mm'}}{2 \pi i} \frac{1}{\om - \bar{\om}} + \text{o}(\frac{1}{\om - \bar{\om}})
\end{align}
so that the spectrum is centered around $\om = \bar{\om}$, but with some non negligible standard deviation. However, the modes $b_{\bar{\om} l m} = \frac{1}{\sqrt{4 \pi \bar{\om}}} Y_m^l e^{- i \bar{\om} \bar{u}}$ are not normalized. In order to get normalized modes, a strategy is to complexify the frequency $\om$ and look at the modes 
\be \label{normalizedmodes}
    b_{\bar{\om} l m}^\eps = \sqrt{\frac{\eps}{\bar{\om} - i \eps}} Y_m^l e^{- i (\bar{\om} - i \eps) \bar{u}} H(\bar{u}), \qquad \eps > 0
\ee
where $\eps$ is small and $H(\bar{u})$ is the Heaviside function. Then, we compute the Klein-Gordon product between two modes \eqref{normalizedmodes} and find
\be
    (b_{\bar{\om} l m}^\eps, b_{\bar{\om}' l' m'}^\eps) = \delta_{ll'} \delta_{mm'} \sqrt{\frac{\bar{\om}' - i \eps}{\bar{\om} - i \eps}} \frac{2 i \eps }{(\bar{\om}' - \bar{\om}) + 2 i \eps} = \left\{ 
    \begin{array}{ll}
        \delta_{ll'} \delta_{mm'} & \mbox{if} \quad \lvert \bar{\om} - \bar{\om}' \lvert \ll \eps \\
        0 & \mbox{if} \quad \lvert \bar{\om} - \bar{\om}' \lvert \gg \eps
    \end{array}
    \right .
\ee
so we have indeed normalized modes in the limit $\eps \rightarrow 0^+$ while 
\begin{align}
    (b_{\bar{\om} l m}^\eps, b_{\om l' m'}) &= 2\delta_{l l'} \delta_{m m'} \int_0^{+ \infty} \sqrt{\frac{\om \eps}{4 \pi (\bar{\om} - i \eps)}}  e^{- i (\bar{\om} - i \eps) \bar{u}} e^{i \om u} \text{d} \bar{u} \\
    \nonumber
    &= \delta_{ll'} \delta_{mm'} \sqrt{\frac{\om}{\pi \bar{\om}}} \frac{\sqrt{\eps}}{(\om - \bar{\om} + i \eps)} + O(\eps^{\f32})
\end{align}
so that, 
\be
    \lvert (b_{\bar{\om} l m}^\eps, b_{\om l' m'}) \lvert^2 = \delta_{ll'} \delta_{mm'} \frac{\om}{\bar{\om}} \frac{1}{\pi} \frac{\eps}{(\om - \bar{\om})^2 + \eps^2} + O(\eps^2) \underset{\eps \rightarrow 0^+}{\longrightarrow} \delta_{ll'} \delta_{m m'} \delta(\om - \bar{\om})
\ee
and we see that the spectral energy distribution of $b_{\bar{\om} l m}^\eps$ is localized around the Killing frequency $\om = \bar{\om}$. Therefore, the normalized outgoing particles emitted from the region $\mathcal{D}_i^{\mathcal{H}_R^-}$ are transmitted to the region $\mathcal{D}_i^\scri$ of $\scri^+_R$ or reflected back towards the black hole horizon $\mathcal{H}^+_L$ with amplitude $t_{\om l}$ and $r_{\om l}$ respectively. Then, we can conclude that the restriction of the Unruh vacuum to the algebra $\mathcal{A}_i^\scri$ of $\scri^+_R$ satisfy the KMS conditions \eqref{KMScondunruhinf} with $\om$ changed into $\bar{\om}$, conjugated to the time \eqref{barutimeui} in the region $\mathcal{D}_i^\scri$, and with $\m_{\bar{\om} l} = \m_{\om l}$, i.e.
\be \label{formalexpressionloc}
    \s^{\mathcal{A}_i^\scri} = \sum_{\{ N_{\bar{\om} l m} \}}\prod_{\bar{\om} > 0, lm} e^{- \beta_H N_{\bar{\om} l m} (\bar{\om} - \m_{\bar{\om} l})} \ket{N_{\bar{\om} l m}} \bra{N_{\bar{\om} l m}}
\ee
Notice that in this case the ladder operators of \eqref{KMScondunruhinf} are changed into the ladder operators associated to the region $\mathcal{D}_i^\scri$, which generate the algebra $\mathcal{A}_i^\scri$.

\subsection{The second law in the Unruh vacuum}

The expression \eqref{formalexpressionloc} allows us to define the one-sided Hamiltonian of the Unruh as
\be \label{onesidunruh}
    K_{\Om_U}^{\mathcal{A}_i^\scri} = 2 \pi \int_{\mathcal{D}_i^\scri} (U - U_i) :T_{UU}:_{\Om_U} \text{d}U \wedge \eps_S - \sum_{lm} \int_{0}^{+ \infty} \frac{\m_{\om l}}{T_H} \bar{n}_{\om l m}  \text{d}\om 
\ee
where we replaced $\bar{\om}$ by $\om$ to simplify the notation, but we should keep in mind that from here $\om$ is not strictly speaking the Killing frequency but the frequency conjugated to the time \eqref{barutimeui} (as we assumed implicitly in the other Sections), and where $\bar{n}_{\omega l m}$ is an unbounded operator defined as 
\begin{equation}
        \label{eq: excess density2}
        \bar{n}_{\om l m} := \underset{\Delta \om \rightarrow 0}{\lim} \sum_{\om, \om + \Delta \om } \frac{N_{\om l m} - \langle N_{\om l m} \rangle_{\Om_U}}{\Delta \om} = n_{\omega l m} - \langle n_{\omega l m} \rangle_{\Omega_U} \, .
    \end{equation}
similarly to the excess density operator \eqref{eq: excess density} defined for the $\kappa_l$-vacuum. In addition, the one-sided modular Hamiltonian \eqref{onesidunruh} satisfies the normalization condition
\be
    \bra{\Om_U} K_{\Om_U}^{\mathcal{A}_i^\scri} \ket{\Om_U} = 0
\ee
and has the same form as the modular Hamiltonian of the $\kappa_l$-vacuum (see \eqref{eq: modhamkappal link unruh}) except that now the chemical potential $\m_{\om l}$ is not linear in the frequency $\om$ (see \eqref{eq: relation mu kappal}) but is now given by \eqref{chemicalpot}. In particular, the modular Hamiltonian \eqref{onesidunruh} is made of a combination of two terms, a geometric one that is giving the variation of internal energy (through the gravitational constraints on $\scri^+_R$) and another one that must be interpreted as work term, and is not equal to the variation of some Noether charge, as it was the case for the soft regularization of the Hartle-Hawking state.

In order to apply the monotonicity of the relative entropy between an arbitrary state $\ket{\Psi}$ in the GNS construction of the Unruh vacuum and the Unruh vacuum itself $\ket{\Om_U}$, between two subalgebras of $\mathcal{A}_0^\scri$, $\mathcal{A}_1^\scri$ and $\mathcal{A}_2^\scri$ with $\mathcal{A}_2^\scri \subset \mathcal{A}_1^\scri$, we must compare the one-sided modular Hamiltonian of the Unruh vacuum attached to these two regions. The arguments of the previous subsection apply whatever the value of $\tilde U_i$ on $\mathcal{H}_R^-$, therefore we conclude that the modular Hamiltonian of the Unruh vacuum associated to the algebra $\mathcal{A}_i^\scri$ has always the form \eqref{onesidunruh} whatever the cut $U_i$ at $\scri_R^+$. Therefore, for a state $\ket{\Psi} \in \mathcal{H}_{\Om_U}$ we have the fundamental inequality
\be
    \Delta S_{\Psi \lvert \Om_U}^{\mathcal{A}_i^\scri} - \Delta \langle K_{\Om_U}^{\mathcal{A}_i^\scri} \rangle_{\Psi} \geq 0
\ee
which has the same form as usual. 

In order to get the charges, we should take into account the flux of energy, entropy and particles in the Unruh vacuum $\ket{\Om_U}$ at $\scri^+_R$. The flux of energy and particles are constant and well known: they are equal to 
\be \label{varianpeuvac}
     \Delta \langle N_{\om l m} \rangle_{\Om_U} = -(u_2 - u_1) \frac{\lvert t_{\om l} \lvert^2}{e^{\beta \om} - 1} \delta \om, \qquad \Delta M_{\Om_U} = -(u_2 - u_1) \sum_{lm} \int_0^{+ \infty} \om \frac{\lvert t_{\om l} \lvert^2}{e^{\beta \om} - 1} \text{d} \om
\ee
where $\langle N_{\om l m} \rangle_{\Om_U}$ is the number of particles with quantum numbers $l$ and $m$ and in the frequency range $(\om, \om + \delta \om)$, with $\delta \om \ll \om$, crossing the section of $\scri^+_R$ between the time $u_1$ and $u_2$, with $u_2 - u_1 \gg \kappa^{-1}$, while $\Delta M_{\Om_U}$ is the variation of the Bondi mass in the Unurh vacuum, given by the total energy flux associated to the particles on the region $(u_1, u_2)$ of $\scri^+_R$. Of course, both terms in \eqref{varianpeuvac} are finite for bounded regions of $\scri^+_R$. In order to deduce the entropy flux at $\scri^+_R$ in the Unruh vacuum, we notice that the flux is stationary so it is associated to a vanishing entropy production. Then, similarly to what we did for the $\kappa_l$-vacuum, in the Unruh vacuum we can consider any bunch of modes with the same angular momentum $l$ and frequency in the range $(\om, \om + \delta \om)$ to be in thermal equilibrium at temperature $T_{\om l}$ given by
\be \label{localtempoml}
    T_{\om l} = \frac{T_H}{1 - \frac{\m_{\om l}}{\om}}
\ee
as we can see from the formal expression \eqref{formalexpressionloc}, where the local temperatures $T_{\om l}$ have been introduced by Page \cite{Page:2004xp} and the effective chemical potentials $\m_{\om l}$ given by \eqref{chemicalpot} in \cite{ARB24}, the relation \eqref{localtempoml} already appearing in that reference. Notice that since there is a infinite number of modes in the range of frequency $(\om, \om + \delta \om)$, no matter how small $\delta \om > 0$ is, we can treat them as an infinite reservoir of modes in thermal equilibrium at the temperature $T_{\om l}$. Therefore, from the Clausius relation we have that
\be
    \delta S_{\om l} = \frac{\delta E_{\om l}}{T_{\om l}}, \qquad \delta E_{\om l} = -(u_2 - u_1) \sum_{m = -l}^{m = + l} \om \frac{\lvert t_{\om l} \lvert^2}{e^{\beta \om} - 1} \delta \om
\ee
and we obtain the total entropy flux by summing the contributions from the local reservoirs at temperature $T_{\om l}$
\be
    \Delta S_{\Om_U} = \sum_{l = 0}^{+ \infty} \int_0^{+ \infty} \delta S_{\om l}
\ee
but from the relation \eqref{localtempoml} we deduce immediately that 
\be \label{equilibriumomlunruhvac}
    \Delta M_{\Om_U} - \sum_{lm} \int_{0}^{+ \infty} \m_{\om l m} \langle \Delta n_{\om l} \rangle_{\Om_U} d \om = T_H \Delta S_{\Om_U},
\ee
and
\begin{equation}
    \langle \Delta n_{\om l m} \rangle_{\Om_U} = \underset{\delta \om \rightarrow 0}{\lim} \frac{\Delta \langle N_{\om l m} \rangle_{\Om_U}}{\delta \om} = -(u_2 - u_1) \frac{\lvert t_{\om l} \lvert^2}{e^{\beta \om} - 1}
\end{equation}
so that for a state (not too far from the vacuum) we can combine \eqref{equilibriumomlunruhvac} with \eqref{varianpeuvac} to get
\be
    \Delta M -  \sum_{lm} \int_{0}^{+ \infty} \m_{\om l} \langle \Delta n_{\om l m} \rangle_{\Psi} d \om - T_H \Delta S_\Psi \leq 0 
\ee
with $\Delta S_\Psi = \Delta S_{\Psi \lvert \Om_U}^{\text{v.N.}, \mathcal{A}_i^\scri} +\Delta S_{\Om_U}$ is the total entropy flux. The same expressions hold upon replacing $\ket{\Psi}$ by $\ket{\bar \Psi} \in \mathscr{H}_{\bar \Omega_U}$ with the total reference state $\ket{\bar \Omega_U} = \ket{\Omega^\mathcal{H}} \otimes \ket{\Omega_U}$.

We notice that we obtain a result that is very similar to the one that we obtained from the $\kappa_l$-vacuum. The only difference is the dependence of the local temperature $T_{\om l}$. However, this difference is crucial, since in the case $T_{\om l} = T_{l}$, it was possible to compute the restriction of the $\kappa_l$-vacuum state on any subalgebra $\mathcal{A}_i^\scri$ of $\scri^+_R$ using the local Mobius symmetries, for each two dimensional chiral CFT associated to the mode $(lm)$. However, we cannot find a time $U_{lm}$ that will make it work for the Unruh vacuum, because the local temperatures depend also on the frequency $\om$. Instead, we used the symmetries of the Unruh vacuum on the past horizon $\mathcal{H}_R^-$ and deduced from a transmission problem a formal expression from the modular Hamiltonian, which is similar to what was done in \cite{ARB24}. However, the proof is not fully satisfactory. In order to get a satisfactory proof, one has to compute the modular Hamiltonian of the Unruh vacuum directly on the subalgebra $\mathcal{A}_i^\scri$ on $\scri^+_R$, instead of using the effective modular Hamiltonians deduced from the formal expressions \eqref{formalexpressionloc}. Consequently, deriving the dual generalized second law directly from the asymptotic algebra $\mathcal{A}_i^\scri$ in the Unruh vacuum remains an open problem.

Nevertheless, the formal expression \eqref{formalexpression} already indicates that the Unruh vacuum is neither a thermal state at $\scri^+_R$ nor even a passive state. It is because from the point of view of an asymptotic observer, the modes are emitted from a bunch of reservoirs with different effective temperatures $T_{\om l}$. Therefore, similarly to what we showed in the previous section, we could use the BLPS engine \cite{brunner2012virtual} in order to lift a load on a energy scale using the Hawking asymptotic outgoing radiation in the Unruh vacuum. 


\subsection{The Kerr black hole}
\label{subsec: Kerr}

\begin{figure}[ht]
        \center
        \includegraphics[width=0.65\textwidth]{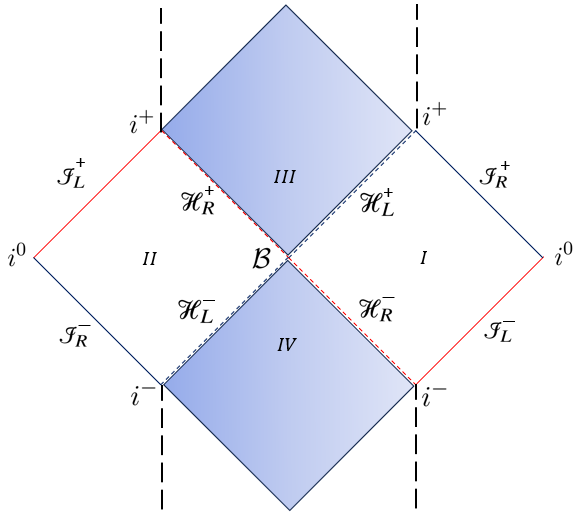}
    \caption{One region of the Penrose diagram of the maximal extension of the Kerr solution. The (anti)-trapped regions (black and white holes) are in shaded blue. In the eternal Kerr solution, the singularities are timelike (vertical dashed black lines). Therefore, contrary to the Schwarzschild solution, one can pile up, on top of each other, an infinite amount of regions like the one depicted in this figure, so that only a fraction of the maximal extension of the Kerr solution is represented here. Notice that as long as one focuses on the exterior regions, the diagram is very similar to the one of an eternal Schwarzschild black hole.}
    \label{fig: penrose eternal Kerr}
    \end{figure}

So far, we have restricted our attention to the Schwarzschild black hole. Realistic astrophysical black holes, however, typically possess nonzero angular momentum. In this case, the spacetime settles into the Kerr solution, a stationary and axisymmetric geometry. This solution admits a one-parameter family of Killing vector fields that become null on the event horizon. At asymptotic infinity, the norm of these Killing fields diverges due to their angular component, reflecting the fact that the black hole rotates with respect to an asymptotic observer with finite angular velocity 
$\Om_H$. Nevertheless, their timelike component can be normalized to (minus) unity at infinity. In this case, one has selected the Killing field
\be \label{killinginftykerr}
    \xi_{\text{Kerr}} = \p_u + \Om_H \p_\phi
\ee
where $u$ is the usual affine coordinate at $\scri^+_I$ (so that $\p_t = \p_u$), while $\p_\phi$ is also a Killing field, being the generator of the rotations in the axisymmetric direction, and its conjugated charge is the angular momentum $L_z$. On both horizons, \eqref{killinginftykerr} becomes
\be \label{killinghorizonkerr}
    \xi_{\text{Kerr}} = \kappa (\tilde{V} \p_{\tilde{V}} - \tilde{U} \p_{\tilde{U}})
\ee
where $\kappa$ is the surface gravity. Therefore, once restricted to the horizon, the Killing field has the same form as the Killing field in Schwarzschild. The maximal extension of the Kerr spacetime, whose Penrose diagram is depicted in Figure \ref{fig: penrose eternal Kerr}, also admits an other universe, and it is overall much richer than the Schwarzschild maximal extension, since the singularities are timelike for example. However, in this work, we focus only on one part of the maximal extension of the solution, the one depicted in Figure \ref{fig: penrose eternal Kerr}. It looks like the maximal extension of the Schwarzschild solution as long as one restricts one's attention to the exterior regions. Therefore, all the arguments given in \cite{RBVilatte251} and summarized in Section \ref{sec: GSL proof} can be generalized to a Kerr background.

\subsubsection{A vacuum state for Kerr}

On both horizons $\mathcal{H}_R$ and $\mathcal{H}_L$ we can define positive frequency solutions with respect to the affine times $\tilde{U}$ and $\tilde{V}$, and a vacuum state $\om$. However, in the Schwarzschild case, we can extend the state $\om$ defined using the algebra on both horizons to the maximally extended spacetime by requiring invariance of the two-point function under the Schwarzschild Killing flow, so that we get the Hartle-Hawking vacuum. However, it is impossible in the case of the Kerr black hole, and this is due to the fact that the Killing field \eqref{killinginftykerr} fails to be timelike in the exterior region \cite{kay1991theorems}.\footnote{In fact, at infinity, the norm of \eqref{killinginftykerr} is infinite because the angular components of the metric diverge in $r^2$.} We could also proceed as in the Schwarzschild case at null infinity, by considering the algebra of observables 
\be
    \mathcal{A}^{\scri_R} = \mathcal{A}^{\scri^+_R} \vee \mathcal{A}^{\scri^-_L}
\ee
and taking the time 
\be
    U = \left\{ 
    \begin{array}{ll}
       e^{\kappa u_+} & \mbox{on} \quad \scri^+_R  \\
        -e^{-\kappa u_-} & \mbox{on} \quad \scri^-_L
    \end{array}
    \right . \, ,
\ee
but the state would not describe the asymptotic behavior of the field's quantum state in a Kerr background for any resonable boundary conditions. First, we would like to break the spherical symmetry, since the Kerr solution is only axisymmetric. In principle, it can be done by choosing for each quantum number $(l,m)$ a positive frequency solution associated to the time coordinate
\be \label{ulmtransf}
    U^{(l,m)} = \left\{ 
    \begin{array}{ll}
       e^{\kappa_{lm} u_+} & \mbox{on} \quad \scri^+_R  \\
        -e^{-\kappa_{lm} u_-} & \mbox{on} \quad \scri^-_L
    \end{array}
    \right .
\ee
that changes for any of the two-dimensional chiral CFT described by the quantum numbers $(l,m)$, as we did in \eqref{ulmsoftdef}. The only difference is that it is now better to take the \textit{real} spherical harmonic basis $\mathcal{Y}_{m}^l$ to decompose the field instead of the complex one.\footnote{
The real spherical harmonics are defined by 
\be
    \mathcal{Y}_{m}^l = \left\{ 
    \begin{array}{ll}
       \frac{Y_m^l + \bar{Y}_{m}^l}{\sqrt{2}} & \mbox{if} \quad m > 0  \\
        Y_m^l & \mbox{if} \quad m = 0 \\
        \frac{Y_m^l - \bar{Y}_{m}^l}{\sqrt{2}i} &\mbox{if} \quad m < 0 \\
    \end{array}
    \right.
\ee
and they are the spherical counterpart of the cosine and sine decomposition of the complex exponential. They also form an orthonormal basis and therefore satisfy the completeness relation. However, they are not eigen-functions of the angular momentum operator $L_z$ (except for $m=0$ of course), but we have instead $L_z \mathcal{Y}_{m}^l = -i m \mathcal{Y}_{-m}^l $ if $m\geq0$ and $L_z \mathcal{Y}_{m}^l = i m \mathcal{Y}_{-m}^l $ otherwise.} 
It is because the transformation \eqref{ulmtransf} mixes the lower operators $\{ a_{\Om l m} \}_{\Om > 0}$ associated with the mode $(l,m)$ with the upper operators $\{ a_{\Om l -m}^\dag \}_{\Om > 0}$ in the spherical harmonic decomposition basis, since $\bar{Y}_m^l = (-1)^m Y_{-m}^l$. Otherwise, everything works exactly as for the $\kappa_l$-vacuum, 
and the quantum field restricted to the algebra $\mathcal{A}_i^\scri$ is described by a thermal state on each sector $(l,m)$ with temperature  $T_{lm} = \frac{\kappa_{lm}}{2 \pi}$, defined with respect to the corresponding local geometric modular Hamiltonian.

Nevertheless, if there is no equivalent to the Hartle-Hawking vacuum state in a Kerr black hole spacetime, the Unruh vacuum still exists. It is because the restriction of the latter on the horizon $\mathcal{H}_R$ is identical to the restriction of the Hartle-Hawking vacuum. Therefore, in the Unruh vacuum, outgoing particles are emitted with a thermal spectrum, at the temperature $T_H = \frac{\kappa}{2 \pi}$ as seen from asymptotic observers, from the past horizon $\mathcal{H}_R^-$. Since \eqref{killinginftykerr} is a Killing field and reduces to \eqref{killinghorizonkerr} on the past horizon (with $\tilde{V} = 0$), then a mode with frequency $\om'$ on the horizon and angular momentum $m$ has a frequency $\om$ (conjugated to the Killing time $u$) as seen by an outgoing observer equal to $\om = \om' + \Om_H m$. Therefore, if all the modes were transmitted perfectly from the past horizon $\mathcal{H}_R^-$ to future infinity $\scri^+_R$, we would have had a thermal spectrum at the Hawking temperature so that for any mode $(\om l m)$ \footnote{Of course, one can also use the KMS conditions to prove that 
\be
    p(N_{\om l m} + 1) = e^{-\beta_H (\om - m \Om_H)} p(N_{\om l m} )   
\ee
and get the complete probability distribution, as usual.
}
\be
    \frac{n_{\om l m} + 1}{n_{\om l m}} = e^{\beta_H (\om - m \Om_H)}
\ee
and from which we could deduce an effective temperature 
\be \label{effectivetemkerr}
    T_{\om m} = \frac{T_{H}}{1 - \frac{m \Om_H}{\om}}
\ee
that is very similar to \eqref{localtempoml} but now with a chemical potential $\m_{m} = m \Om_H$.\footnote{The quantity $m \Om_H$ has already been associated to a chemical potential in the literature, see for instance \cite{navarro2005modeling}.} However, while looking closely, there are several important differences with the formula \eqref{localtempoml}. The most important difference is that the temperature \eqref{effectivetemkerr} can be negative, while it was impossible for \eqref{localtempoml} since the chemical potential was negative. It happens when $\om - m \Om_H < 0$. In thermodynamics, a negative temperature is a feature of instability and indicates that work can be extracted from the system. For instance, the virtual qubit of the BLPS engine \cite{brunner2012virtual} studied in Section \ref{sec: work extraction} has a negative temperature in the case of a population inversion. The possibility of extracting work from the Kerr black hole is well-known and translates classically into the Penrose process \cite{penrose1971extraction}, in which the mass of the black hole can decrease if the mass loss is counterbalanced by an higher loss in angular momentum, so that the variation of the black hole area remains positive in the process, which, by conservation of the Killing energy and the angular momentum, means for the escaping particle that $E - \Om_H J_z < 0$ (which is the classical counterpart of $\om - \Om_H m < 0$). Likewise, \textit{superradiance} is a quantum manifestation of the instability of a Kerr black hole and happens precisely for the modes satisfying $\om - \Om_H m < 0$. It can be seen from the fact that in a Kerr spacetime the formula \eqref{rtconseq} becomes
\be \label{gammakerr}
    \Gamma_{\om l m} = \left(1 - \frac{m \Om_H}{\om}\right) \lvert t_{\om l m} \lvert^2 = 1 - \lvert r_{\om l m} \lvert^2
\ee
so that $\lvert r_{\om l m} \lvert > 1$ if  $\om - \Om_H m < 0$ and therefore there is a spontaneous emission of particles from the past horizon. Notice that even if  $\Gamma_{\om l m}$ can be negative, we have $\G_{\om l m} < 1$.

Therefore, from the above discussion, one can compute the restriction of the Unruh vacuum of a Kerr black hole on $\scri^+_R$ by using the same strategy as for the Schwarzschild black hole, i.e. by studying a transmission/reflexion process, except that now 
we have to replace $\om$ by $ \om - \Om_H m$ and the transmission coefficient $ \lvert t_{\om l m} \lvert^2$ appearing in the chemical potential by $\Gamma_{\om l m}$ \eqref{gammakerr}. Then, we can conclude that the formal one-sided modular Hamiltonian on $\scri^+_R$ for the Unruh vacuum in a Kerr black hole is given by 
\be \label{formalexpressionlockerr}
    \s^{\mathcal{A}_0^\scri} = \sum_{\{ N_{\om l m} \}}\prod_{\om > 0, lm} e^{- \beta_H N_{\om l m} (\om - m \Om_H - \m_{\om l m})} \ket{N_{\om l m}} \bra{N_{\om l m}}
\ee
with chemical potentials \cite{ARB24}
\be \label{chemicalpotkerr}
    \m_{\om l m} = T_H \ln{ \frac{\Gamma_{\om l m}}{1 - (1 - \Gamma_{\om l m})e^{- \beta(\om - m \Om_H)}}}
\ee
and effective temperature 
\be \label{tomlkerr}
    T_{\om l m} = \frac{T_H}{1 - \frac{m \Om_H + \m_{\om l m}}{\om}} \, .
\ee
One can show that $\om - m \Om_H - \m_{\om l m} \geq 0$ \cite{ARB24} as one can write
\begin{equation}
    \label{proofstep1}
    \om - m \Om_H - \m_{\om l m} =  \om - m \Om_H - T_H \ln{\lvert \G_{\om l m} \lvert} + T_H \ln{\lvert 1 -(1  - \G_{\om l m})e^{- \beta_H(\om - m \Om_H)}\lvert}
\end{equation}
and then decompose the proof into two sub-cases. When $\om - m \Om_H > 0$ we have
\begin{align}
    \label{proofcase1}
    \lvert \G_{\om l m} \lvert &= \G_{\om l m} \\ 
    \lvert 1 - (1 - \G_{\om l m})e^{- \beta_H(\om - m \Om_H)}\lvert &= 1 - (1 - \G_{\om l m})e^{- \beta_H(\om - m \Om_H)} \\
    &= 1 - e^{- \beta_H(\om - m \Om_H)} + \G_{\om l m} e^{-\beta_H(\om - \Om_H m)} \\
    &\geq \G_{\om l m} e^{-\beta_H(\om - \Om_H m)} 
\end{align}
and we conclude from the fact that the logarithm is an increasing function. When $\om - m \Om_H < 0$ we have instead
\begin{align}
    \label{proofcase 2}
    \lvert \G_{\om l m} \lvert &= -\G_{\om l m} \\
    \lvert 1 - (1 - \G_{\om l m})e^{- \beta_H(\om - m \Om_H)}\lvert &= (1 - \G_{\om l m})e^{- \beta_H(\om - m \Om_H)} - 1 \\
    &= e^{- \beta_H(\om - m \Om_H)} - 1 + \lvert \G_{\om l m} \lvert e^{-\beta_H(\om - \Om_H m)} \\
    &\geq \lvert \G_{\om l m} \lvert e^{-\beta_H(\om - \Om_H m)}
\end{align}
and we also conclude from the fact that the logarithm is an increasing function. Therefore the effective temperature \eqref{tomlkerr} is always positive. Similarly, it can be proven that \eqref{chemicalpotkerr} is always negative, as in the Schwarzschild case.\footnote{The proof goes along the same line as for the positivity of the temperature.}

Now, if we want to restrict the Unruh vacuum to the algebra $\mathcal{A}_i^\scri$ associated to the region $\mathcal{D}_i^\scri$, we can proceed for Kerr exactly as we did for Schwarzschild in the previous section. We use the symmetries of the Unruh vacuum on the past horizon $\mathcal{H}_R$ to have a KMS state on some subregion $(\tilde U_i, 0) \times S^2_\mathcal{H}$ of the past horizon $\mathcal{H}_R^-$ with respect to the modular Hamiltonian \eqref{modhamiltonianuihorr}, and conclude that the modes now have a frequency $\bar \omega$ associated to the correct time $\bar u$ defined in \eqref{modaffinetimeu0draft2}. Similar computations than in the Schwarzschild case show that the distribution of the frequencies $\bar \omega$ is highly picked around the Killing frequency $\omega$ so that the outgoing modes are still transmitted using the same gray-body factor $\Gamma_{\om l m}$. Hence, the formal one sided modular operator associated to the algebra $\mathcal{A}_i^\scri$ is given by 
\be \label{formalexpressionlockerr2}
    \s^{\mathcal{A}_i^\scri} = \sum_{\{ N_{\bar{\om} l m} \}}\prod_{\bar{\om} > 0, lm} e^{- \beta_H N_{\bar{\om} l m} (\bar{\om} - m \Om_H - \m_{\bar{\om} l m})} \ket{N_{\bar{\om} l m}} \bra{N_{\bar{\om} l m}}
\ee
where $\bar{\om}$ is conjugated to the time $\bar{u} = u + \kappa^{-1} \ln{(1 - e^{- \kappa(u - u_i)})}$, as in the Schwarzschild case. 

\subsubsection{The second law in a Kerr background}

To avoid clutter we set $\bar \omega = \omega$ in this paragraph, but one should remember that they are strictly speaking different. If we consider that the formula for the one-sided modular operator associated with the algebra $\mathcal{A}_i^\scri$ is given by \eqref{formalexpressionlockerr2}, then, the one-sided modular Hamiltonian of the Unruh vacuum in the Kerr background is given by \footnote{We use \eqref{eq: excess density2} for defining for the various objects appearing in \eqref{modhamkerr}.}
\begin{align}
\label{modhamkerr}
     K_{\Om_U}^{\mathcal{A}_i^\scri} = 2 \pi \int_{\mathcal{D}_i^\scri} (U - U_i) :T_{UU}:_{\Om_U} \text{d}U \w \eps_S &- \sum_{lm} \int_{0}^{+ \infty} \frac{\m_{\om l m}}{T_H} (n_{\om l m} -\langle n_{\om l m} \rangle_{\Om_U}) \text{d} \om \\
     \nonumber
     &- \frac{\Om_H}{T_H} \sum_{lm} \int_{0}^{+ \infty} m (n_{\om l m}  - \langle n_{\om l m} \rangle_{\Om_U}) \text{d} \om
\end{align}
that we normalized so that 
\be
    \bra{\Om_U} K_{\Om_U}^{\mathcal{A}_i^\scri} \ket{\Om_U} = 0
\ee
as always. Compared to \eqref{onesidunruh}, the only new additional term in \eqref{modhamkerr} is the (renormalized) angular momentum flux 
\begin{align} 
    L_z^i - \langle L_z^i \rangle_{\Om_U} &= \sum_{lm} \int_{0}^{+ \infty} m (n_{\om l m}  - \langle n_{\om l m} \rangle_{\Om_U}) \text{d} \om \nn \\
    &= \int_{\mathcal{D}_i^\scri} (\p_u \phi \p_\phi \phi - \langle \p_u \phi \p_\phi \phi \rangle_{\Om_U}) \text{d}u \wedge \eps_S
\end{align}
where we used the fact that $\p_\phi Y_m^l = i m Y_m^l$. Then, in order to get the charge variation, we have to compute the flux of energy, angular momentum and number of particles in the Unruh vacuum in a Kerr background. The fluxes between two cross sections $u = u_1$ and $u = u_2$ of $\scri^+_R$ are constant and given by
\begin{align} \label{varianpeuvackerr}
     \Delta \langle N_{\om l m} \rangle_{\Om_U} &= -(u_2 - u_1) \frac{\Gamma_{\om l m}}{e^{\beta (\om - \Om_H m)} - 1} \delta \om, \qquad \Delta M_{\Om_U} = -(u_2 - u_1) \sum_{lm} \int_0^{+ \infty} \om \frac{\Gamma_{\om l m}}{e^{\beta (\om - \Om_H m)} - 1} \text{d} \om \nn \\
     \Delta J_{\Om_U} &= -(u_2 - u_1) \sum_{lm} \int_0^{+ \infty} m \frac{\Gamma_{\om l m}}{e^{\beta (\om - \Om_H m)} - 1} \text{d} \om \, .
\end{align}
Then, we deduce the entropy flux in the Unruh vacuum in a manner exactly similar to the Schwarzschild case, except that for Kerr we have to take into account the angular momentum flux. Again, we consider a subset of modes with quantum numbers $(l,m)$ and with frequency in the range $(\om + \delta \om)$ and treat them as a reservoir with local temperature $T_{\om l m}$. Recall that since the modes are labeled in frequency by a continuous parameter $\om$, there is an infinite number of modes in any of our local reservoirs, so that we can consider them as infinite. Therefore, the local Clausius relations in the stationary Unruh vacuum (no entropy production at all) tell us that 
\be
    \delta S_{\om l m} = \frac{\delta E_{\om l m}}{T_{\om l m}}, \qquad \delta E_{\om l m} = -(u_2 - u_1) \om \frac{\G_{\om l m}}{e^{\beta (\om- \Om_H \om)} - 1} \delta \om 
\ee
so that
\be
    \Delta S_{\Om_U} = \sum_{lm} \int_0^{+ \infty} \delta S_{\om l m}
\ee
and from \eqref{tomlkerr} we deduce that 
\be \label{equilibriumomlunruhvackerr}
    \Delta M_{\Om_U} - \Om_H \Delta J_{\Om_U} - \sum_{lm} \int_{0}^{+ \infty} \m_{\om l m} \langle \Delta n_{\om l m} \rangle_{\Om_U} d \om = T_H \Delta S_{\Om_U}.
\ee
Then, we obtain the second law from the monotonicity of relative entropy for a restriction of algebra between a state $\ket{\Psi} \in \mathscr{H}_{\Om_U}$ and the Unruh vacuum $\ket{\Om_U}$
\be
    \Delta S_{\Psi \lvert \Om_U}^{\text{v.N.}, \mathcal{A}_i^{\scri}} - \Delta \langle K_{\Om_U}^{\mathcal{A}_i^\scri} \rangle_{\Psi} \geq 0
\ee
combined with the relation \eqref{equilibriumomlunruhvackerr} giving the charge fluxes in the Unruh vacuum. In the end, we find that
\be \label{thermopotkerr}
    \Delta M - \Om_H \Delta J -  \sum_{lm} \int_{0}^{+ \infty} \m_{\om l m} \langle \Delta n_{\om l m} \rangle_{\Psi} d \om - T_H \Delta S_\Psi \leq 0 
\ee
where $J$ is the geometric Penrose angular momentum charge \cite{penrose1982quasi} at $\scri^+_R$, and whose variation is equal to the flux of angular momentum on the corresponding portion of $\scri^+_R$ thanks to the semi-classical Einstein equations. Hence we found another thermodynamic potential associated to the Unruh vacuum in a Kerr background, which reduces to the one we found in the $\kappa_l$-vacuum in a Schwarzschild background for $\Om_H = 0$. In addition, we should interpret the angular momentum term and the particle flux term as work, as it can be seen from the local temperature \eqref{tomlkerr}. Indeed, we can run the BLPS engine (or another autonomous engine) and the maximal amount of extractable work should be given by Carnot's efficiency in the reversible case $\frac{Q_1}{T_1} = \frac{\om}{T_{\om l m}} = \frac{\om'}{T_H} =  \frac{Q_2}{T_2}$, i.e.
\be
    \lvert W_{\text{max}} \lvert = \eta_C Q_2 =  (1 - \frac{T_{\om l m}}{T_H}) \om' = (\frac{T_H}{T_{\om l m}} - 1) \om = - \m_{\om l m} - m \Om_H.
\ee
However, notice that there is a difference between the two work terms appearing in \eqref{thermopotkerr}. Indeed, the angular momentum flux gives a geometric contribution thanks to the semiclassical Einstein's equations at $\scri^+_R$, while it is not the case for the particle flux.


\section{Conclusion and outlooks}
\label{sec: conclusion}

In writing this two-part series (together with \cite{RBVilatte251}), our aim was to address readers from diverse backgrounds with an interest in thermodynamics, and to clarify the precise connections between concepts from quantum thermodynamics and open quantum systems on the one hand, and recent developments in quantum field theory and gravity on the other hand, with particular emphasis on thermodynamics on null hypersurfaces. We hope that the results presented here will be of interest to multiple communities and contribute to building bridges between subfields that share a common focus on thermodynamic principles.

In the present paper, we place particular emphasis on the close connection between the quantization of a free field on a non-expanding null hypersurface, such as $\scri_R$, and the emergence of Markovian dynamics in open quantum systems. From this perspective, black hole thermodynamics appears strikingly similar to quantum thermodynamics in the weak-coupling regime. This analogy naturally motivates the use of additional tools and concepts from quantum thermodynamics and quantum stochastic thermodynamics to gain further insights into black hole physics. One important outcome of this viewpoint is the clarification of the relation between the multiplicity of vacuum states that can be defined in a black hole background and the one-to-one correspondence we have emphasized between the choice of late-time boundary conditions compatible with these vacuum states and the induced thermodynamic potentials. This correspondence also underlies the effective decomposition between work and heat fluxes provided by the modular Hamiltonian.

This framework may therefore be used to extend proofs of the generalized second law to more general settings. Wall’s proof of the generalized second law \cite{Wall:2011hj}, for instance, relies on a Hilbert space obtained as a representation of the horizon algebra built from the Hartle–Hawking vacuum state, which is compatible with an expansion that vanishes at late times (see below \eqref{differenceareasde1draft2}). However, realistic black holes evaporate and do not relax toward equilibrium states, so that this condition is not satisfied. As a consequence, the Hartle–Hawking and Unruh vacuum states exhibit different infrared behaviors and do not belong to the same Hilbert space. A closely related discussion appears in \cite{Faulkner:2024gst}. To overcome these difficulties, one must either study the second law for a broader class of late-time boundary conditions—corresponding to a different, more physically realistic Hilbert space, as we have done at null infinity in the present work and in \cite{RBVilatte251}—or formulate the problem directly on an open region with compact closure, so that the specific choice of Hadamard vacuum state becomes irrelevant.

Another promising direction suggested by the above considerations concerns the role of possible non-Markovian effects in the formulation of the second law in semiclassical gravity. Throughout this paper, we have seen that the thermodynamics of quantum fields on non-expanding null hypersurfaces is naturally associated with Markovian dynamics in open quantum systems, since the presence of a causal horizon prevents any backflow of information. By contrast, one may investigate the monotonicity of thermodynamic potentials—such as the generalized entropy—on non-causal horizons, including dynamical horizons \cite{Ashtekar:2004cn} or stretched horizons \cite{price1986membrane, Freidel:2024emv}. This question is also closely related to recent work on black hole dynamical entropies \cite{Rignon-Bret:2023fjq, Hollands:2024vbe, Visser:2024pwz}. Since neither dynamical horizons nor stretched horizons are null hypersurfaces in general, information backflow is no longer a priori excluded, and non-Markovian effects may become relevant for the formulation of the second law. In such a non-Markovian regime, the second law cannot be expressed as a local, monotonic entropy-production rate. Instead, it must be formulated as a global entropy balance or, equivalently, as a relative-entropy inequality, in which entropy production is governed by the buildup of system–environment correlations rather than by irreversible dissipation alone \cite{esposito2010entropy, de2017dynamics, marcantoni2017entropy, popovic2018entropy}. It would therefore be very interesting to understand how the generalized second law is realized in this setting, and whether further parallels with quantum thermodynamics can be established.

Finally, from a complementary perspective, it may be particularly fruitful to study alternative thermodynamic potentials—such as the generalized free energy—rather than focusing exclusively on the generalized entropy itself. From the viewpoint of asymptotic observers, the black hole area is not a directly measurable quantity, whereas asymptotic charges such as the Bondi or ADM mass are observable \cite{ARB24}. In this sense, free energy (or related thermodynamic potentials), rather than entropy alone, becomes the more natural thermodynamic quantity, as already emphasized in \cite{ARB24} and in the present series of papers. Notably, both the black hole area and the mass are geometric quantities that arise as Noether charges: the former is associated with Killing fields at the horizon, while the latter is tied to asymptotic symmetries at infinity \cite{wald1993black, iyer1994some}. If spacelike slices are compact, as in de Sitter space, such boundary charges vanish altogether. This observation suggests that entropy and energy share a common conceptual origin in covariant theories of gravity. Both emerge as boundary charges associated with diffeomorphism symmetries and differ only in the geometric nature of the boundary on which they are defined. From this perspective, there is no deeper conceptual difficulty in understanding entropy in gravity than in understanding energy or free energy. Accordingly, investigating alternative thermodynamic potentials may provide new insights into the microscopic behavior of gravity and the underlying thermodynamic structure of spacetime.

\section*{Acknowledgments}

ARB would like to thank Mohamed Boubakour, Luca Ciambelli, Cyril Elouard, Anthony Speranza, Simone Speziale and Aron Wall for stimulating discussions on this topic. MV thanks Cyril Elouard for discussions during his visit in Nancy. ARB is also grateful to the Julian Schwinger Foundation for financial support at the 2025 Peyresq Spacetime Meeting, where useful discussions influenced this work. The work of ARB is funded by the European Union. Views and opinions expressed are however those of the author(s) only and do not necessarily reflect those of the European Union or the European Research Council. Neither the European Union nor the granting authority can be held responsible for them. This work is supported by ERC grant QARNOT, project number 101163469. The work of MV is supported by the Fonds de la Recherche Scientifique -- FNRS under the Grant No. T.0047.24. We thank each other institutions for hosting and financial support during visits at the last stages of this work.

\appendix


\bibliographystyle{style}
\bibliography{bibliographe.bib}

\end{document}